\documentclass[final,3p,twocolumn]{elsarticle}

\usepackage{amsmath}
\usepackage{txfonts}
\usepackage[utf8]{inputenc}
\usepackage{graphicx}
\usepackage{makecell} 

\usepackage{hyperref,xcolor}
\usepackage[T1]{fontenc} 
\usepackage{graphicx}
\usepackage{ulem}

\usepackage{amssymb} 
\usepackage{tabularx} 
\usepackage{gensymb}
\usepackage{appendix}
\usepackage{enumerate}   
\usepackage{bm}
\usepackage{setspace}
\usepackage{textcomp}

\usepackage{subcaption}
\usepackage{caption}

\journal{Astroparticle Physics}

\begin{document}

\begin{frontmatter}

\title{Radio Morphing: Fast computation of inclined air shower radio emission\\ }

\author[a,b]{Simon Chiche}
\ead{simon.chiche@ulb.be}
\author[c,b]{Olivier Martineau-Huynh}
\author[d,e,b]{Matias Tueros}
\author[f]{Krijn D. de Vries}

\address[a]{Inter-University Institute For High Energies (IIHE), Universit\'{e} libre de Bruxelles (ULB), Boulevard du Triomphe 2, 1050 Brussels, Belgium}
\address[b]{Sorbonne Universit\'{e}, CNRS, UMR 7095, Institut d'Astrophysique de Paris, 98 bis bd Arago, 75014 Paris, France}
\address[c]{Sorbonne Université, Université Paris Diderot, Sorbonne Paris Cité, CNRS,Laboratoire de Physique Nucléaire et de Hautes Energies (LPNHE), Paris, France}
\address[d]{IFLP - CCT La Plata - CONICET, Diagonal 113 y Calle 63, La Plata, Argentina}
\address[e]{Depto. de Fisica, Fac. de Cs. Ex., Universidad Nacional de La Plata,  Casilla de Coreo 67 (1900) La Plata, Argentina}
\address[f]{Vrije Universiteit Brussel, Physics Department, Pleinlaan 2, 1050 Brussels, Belgium}

\date{Received $\langle$date$\rangle$ / Accepted $\langle$date$\rangle$}

\begin{abstract}

The preparation of the next-generation of large-scale radio experiments requires running a large number of simulations to explore multiple detector configurations over vast areas and develop novel methods for the reconstruction of air shower parameters.  While Monte Carlo simulations are accurate and reliable tools, they are too computationally expensive to explore the full parameter space of these new detectors within a reasonable timescale, and faster and more efficient methods are needed. We introduce a new version of Radio Morphing, a semi-analytical tool designed to simulate the radio emission of any cosmic-ray induced air shower with zenith angle $\theta>60^{\circ}$, at any desired antenna position, from the simulation data of a few reference showers at given positions. This version incorporate refined scaling laws of the radio emission with the shower zenith angle, a novel interpolation method, the implementation of the charge-excess mechanism and the possibility to enable shower-to-shower fluctuations. We present the latest performances of Radio Morphing, tuned for a GRAND-like detector,  which now provides an estimation of air shower radio signals four orders of magnitude faster than standard Monte Carlo simulations, while keeping an accuracy on the peak amplitude better than $17\%$ on unfiltered traces, better than $13\%$ in the [50 - 200] MHz band, and  below $\sim 10\%$ in the [30 - 80] MHz band.

\end{abstract}

\begin{keyword}
Radio-detection, High-energy astroparticles, Air-shower simulations
\vspace{0.5cm}
\noindent

\end{keyword}

\end{frontmatter}


\newpage

\newcommand{\red}{\color{red}}


 \begin{figure*}[tb]
\centering 
\includegraphics[width=0.99\linewidth]{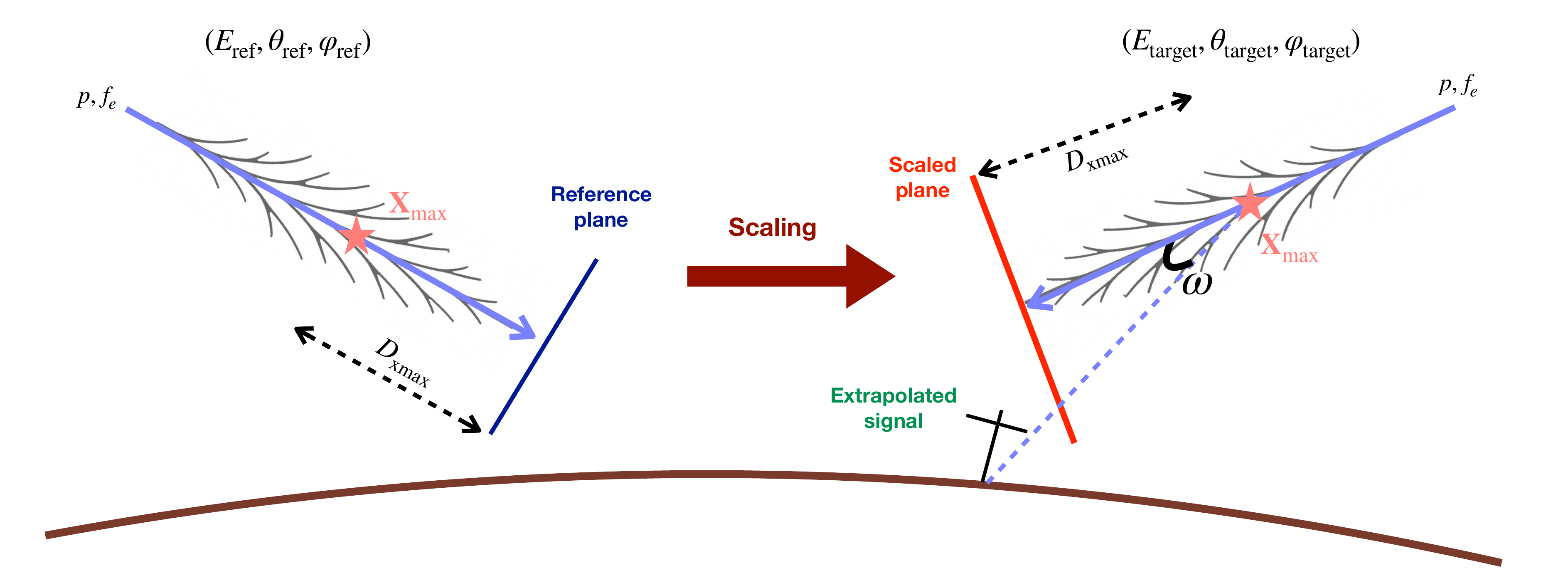}
\caption{Sketch of the Radio Morphing procedure. The electric field from a reference ZHAireS simulation with parameters [$E_{\rm ref}, \theta_{\rm ref}, \varphi_{\rm ref}, D_{\rm xmax}$] (blue ``reference plane'') is scaled towards a shower with [$E_{\rm target}, \theta_{\rm target}, \varphi_{\rm target}, D_{\rm xmax}$] (red ``scaled plane''). The resulting signal is then interpolated within the plane and extrapolated along the shower axis to infer the radio-emission at any position.}\label{fig:sketch_rm}
\end{figure*}

\section{Introduction}


Radio detection is now a mature, well-established technique, used by a wide range of experiments for cosmic ray or neutrino-induced air shower detection to unveil the high-energy Universe. Large-scale radio experiments such as GRAND~\cite{GRAND2020SCPMA..6319501A}, BEACON~\cite{Beacon}, SKA~\cite{SKA2025PhRvD.112b3017C}, or the radio upgrade of the Pierre Auger Observatory~\cite{HuegeAERA2019EPJWC.21005011H} are now being built to reach unprecedented sensitivity, field of view, and angular resolution probing the cosmic particle flux at the highest energies. This goal requires  to run a large number of air-shower simulations to explore different antenna configurations, test reconstruction methods, and evaluate the detector performances. While Monte Carlo simulations such as ZHAireS~\cite{ZHAireS, AlvarezZHS2012APh....35..325A} and Coreas~\cite{CORSIKA1998cmcc.book.....H,HuegeCOREAS2013AIPC.1535..128H} provide accurate simulation results, these are too computationally intensive to explore all the parameter space with large-scale experiments. On the other hand, macroscopic approaches such as MGMR~\cite{Scholten2008APh....29...94S, Scholten:2017tcr} are much faster but limited by our incomplete knowledge of the physical processes, particularly when looking for showers with very inclined arrival directions.  This motivates the need to develop fast and accurate estimations of the radio signal. Already significant efforts have been made towards this direction, with template synthesis for vertical and mildly inclined showers~\cite{Mitja2026APh...17503182D, Tueros2021JInst..16P2031T}. On the other hand, reconstruction of inclined showers remain challenging because of asymmetry effects and rapid changes in the emission with the shower zenith angle~\cite{Schlueter2020EPJC...80..643S, Guelfand2025APh...17103120G, chiche2024PhRvL.132w1001C}.  We introduce here a new version of Radio Morphing, a semi-analytical tool designed to estimate the radio emission from any air shower with zenith angle above $60^{\circ}$, at any position, from the reference data of a few simulated showers at given positions. With Radio Morphing, we aim to globally reproduce the results of standard Monte Carlo simulations for inclined air showers, with relative errors on the electric field amplitude below $20\%$, corresponding to the typical calibration uncertainties expected for radio experiments. The current version incorporates shower-to-shower fluctuations, refined scaling laws, interpolation method, and reference library compared to the version of~\cite{Zilles2020APh...114...10Z}.  
We present the current Radio Morphing method and its implementation in Sections~\ref{Sec:Method} and~\ref{sec:implementation} respectively, and show the latest Radio Morphing performances in Section~\ref{sec:performances}.

\section{The Radio Morphing method}\label{Sec:Method}

The Radio Morphing method is related to the principle of the universality of air showers, which assumes that the radio emission, supposed point-like, can be fully inferred by characterizing the shower at its maximum development $X_{\rm max}$~\cite{Universality2006APh....24..484G}. The point-like assumption becomes well justified for inclined air showers $(\theta \gtrsim 60^{\circ})$ for which the ground to $X_{\rm max}$ distance extends over tens to hundreds of kilometers and is therefore much larger than the typical size of the emitting region. Under this assumption, the variations in the radio emission between different showers can directly be modeled by accounting for the variations at $X_{\rm max}$. While this  simplifies the modeling of the radio emission, it also restricts the applicability of Radio Morphing to inclined air showers $(\theta \gtrsim 60^{\circ})$. 

\subsection{First principles of Radio Morphing}
The Radio Morphing procedure can be divided into two steps illustrated in Fig.~\ref{fig:sketch_rm}: 

 (1) First, the {\it scaling} procedure: We select a reference shower, simulated with ZHAireS, from our library (introduced in Section~\ref{sec:lib}), for which the electric field was simulated at sampled antenna positions in planes perpendicular to the shower axis located at various distances from $X_{\rm max}$, as illustrated by the blue ``reference'' plane in Fig.~\ref{fig:sketch_rm}. The electric field traces of this shower are then scaled using simple laws based on physical principles to infer the radio emission from a target shower with different primary composition, energy, and arrival direction, at sampled antenna positions located at the same $X_{\rm max}$ distance $D_{\rm xmax}$ and aperture angle $\omega$ as the reference ones (scaled plane in Fig.~\ref{fig:sketch_rm}). 
 
 (2) The {\it interpolation} procedure: we then use the electric field traces at the scaled positions to interpolate the radio signal at any position within the plane. Eventually, we  extrapolate the radio emission at any given position in space along the shower axis, indicated for example by the green ``extrapolated'' signal. 
 
The Radio Morphing procedure is performed on the full-band electric field traces, to infer general results that can then be filtered in the desired frequency range, similarly to Monte Carlo simulations.



\section{Radio Morphing implementation}~\label{sec:implementation}

To morph a reference shower with primary energy, arrival direction and mass number $(E_r$, $\varphi_r$, $\theta_r, A_r)$ into a target shower with $(E_t$, $\varphi_t$, $\theta_t, A_t)$, we need to account for the variation of the radio emission with the shower parameters. For this, we can rely on scaling laws, mainly based on the physical properties of the shower emission. The current Radio Morphing implementation follows similar principles to the version of~\cite{Zilles2020APh...114...10Z}, but with several refinements to improve its efficiency and range of validity. For completeness, we detail the full scaling procedure and highlight the specific improvements when appropriate.

\subsection{Reference library and plane selection}~\label{sec:lib}

\begin{table}[b]
\centering
\begin{tabular}{|c|c|}
\hline
\textbf{Energy [EeV]} \rule{0pt}{2.5ex} & 3.98 \\
\hline
\textbf{Zenith [°]} & \makecell[l]{
65.5; 67.8; 69.8; 74.8; 76.1; 77.4; \\
78.5; 79.5; 80.4; 82.7; 83.4; 84.5; \\
85.0; 86.2; 86.5; 86.8
} \\
\hline
\textbf{Azimuth [°]} & 90 \\
\hline
\end{tabular}
\caption{Radio Morphing reference library. The library consists of 16 individual simulations run with ZHAireS, corresponding to one primary particle energy, one azimuth angle, and sixteen zenith angles.}
\label{table:reference}
\end{table}
 Radio Morphing requires first to build a set of reference simulations, that we simulated with ZHAireS and that can be used to scale the radio emission towards the target showers. The initial Radio Morphing implementation was using only one reference shower, but we chose to use multiple showers to reduce the error introduced by the scaling procedure and improve the efficiency of the method. These showers must be chosen wisely to sample the parameter space at limited computing costs. For our reference library, we simulated 16 ZHAireS iron-induced showers (Table~\ref{table:reference}). Iron primaries were favored compared to lighter ones, because of their reduced shower-to-shower fluctuations, but these showers can then be used to model showers for any other cosmic ray primary. All our reference showers were chosen with a primary energy  of $3.98\, \rm EeV$ and  an azimuth angle $\varphi = 90^{\circ}$ (shower propagating towards the East). We considered 16 zenith bins linear in $1/\log_{10}(\cos{\theta})$, as shown in Table~\ref{table:reference}. This dense sampling in zenith angle enables the method to capture the rapid variations of the radio emission with the shower inclination, without implementing an in-depth modeling of complex effects~\cite{Schlueter2020EPJC...80..643S,Guelfand2025APh...17103120G, chiche2024PhRvL.132w1001C}, by selecting the closest reference shower in zenith for each target event generated with Radio Morphing. Computation points were placed in planes perpendicular to the shower axis, at various distances between $X_{\rm max}$ and a ground altitude of $1000\,$ meters above sea level. Each plane consists of a star-shape layout with 193 antennas, but the number of planes used typically increases with zenith angle to account for the larger ground-$X_{\rm max}$ distance at higher inclination, which is another improvement compared to~\cite{Zilles2020APh...114...10Z}. For all our simulations we used SIBYLL23d as hadronic model, a typical magnetic field of Dunhuang (China) with amplitude $B=56\, \rm \mu T$, and inclination $i = 60.8^{\circ}$, Linsley 5-layers atmospheric model~\cite{Linsley} and a relative thinning of $10^{-5}$.

When generating signals with Radio Morphing, the user must specify a target primary energy, zenith and azimuth angle $(E_{t},\, \theta_t, \, \varphi_t)$. We then choose, in our library, the reference ZHAireS simulation with the closest zenith angle $\theta_{\rm ref}$. The user must also specify a coordinate list of target antenna positions ($x_t$, $y_t$, $z_t$) in the AIRES reference frame, at which the radio signal will be computed. Hence, for each target antenna, we take within the selected ZHAireS simulation, the reference plane with $X_{\rm max}$ distance $D_{\rm plane}$ that is the closest to the $X_{\rm max}$-antenna distance $d_{\rm ant}$. This procedure is done to reduce as much as possible the error related to the extrapolation of the signal.

\begin{figure*}[tb]
\includegraphics[width=0.49\linewidth]{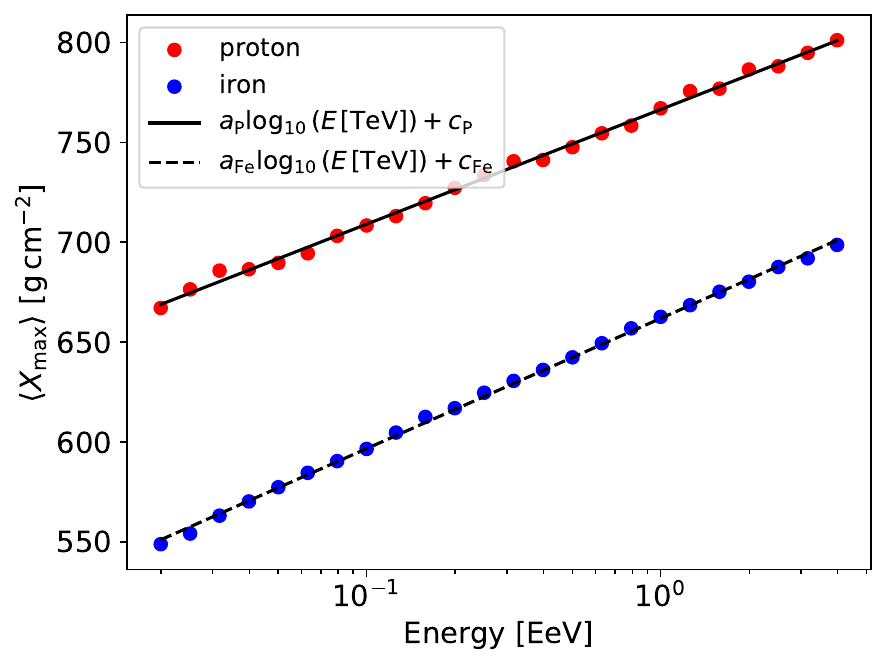}\hfill
\includegraphics[width=0.49\linewidth]{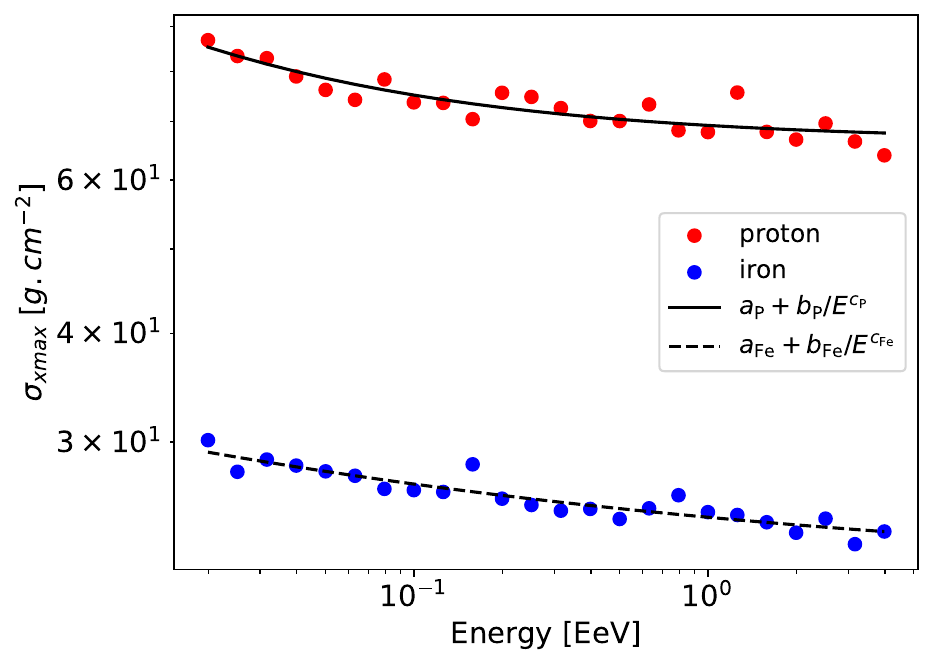}
\caption{Mean ({\it left)} and standard deviations ({\it right)} of $X_{\rm max}$, for proton (red dots) and iron (blue dots) induced air showers, from a set of $11\, 000$ ZHAireS simulations using SYBILL23d as hadronic model. The mean $X_{\rm max}$ values are fitted with a function following $\langle X_{\rm max} \rangle=a\log_{10}(E[{\rm TeV}])+c$, while the standard deviations are parametrized with $\sigma_{\rm xmax} = a + b/E^c$, with the full black lines corresponding to proton primaries and the dashed back lines to iron primaries.}\label{fig:XmaxParam}
\end{figure*}

\subsection{Primary composition}~\label{sec:primary_scaling}
For cosmic ray showers, changes in the primary nature mainly affect the radio emission through a modification of the $X_{\rm max}$ position, since showers induced by light primaries develop deeper in the atmosphere than those induced by heavy nuclei. In this new version of Radio Morphing, we  model these effects by parametrizing the mean $X_{\rm max}$ grammage of proton and iron induced air showers, with $\sim 11\,000$ ZHAireS simulations, using SYBILL23d as hadronic model, following
\begin{equation}
    \langle X_{\rm max} \rangle = a\log_{10}{E[\rm TeV]} +c \ , \label{eq:XmaxParam}
\end{equation}

\begin{figure*}[tb] 
    \centering

    \begin{subfigure}[t]{0.45\textwidth}
        \includegraphics[width=\linewidth]{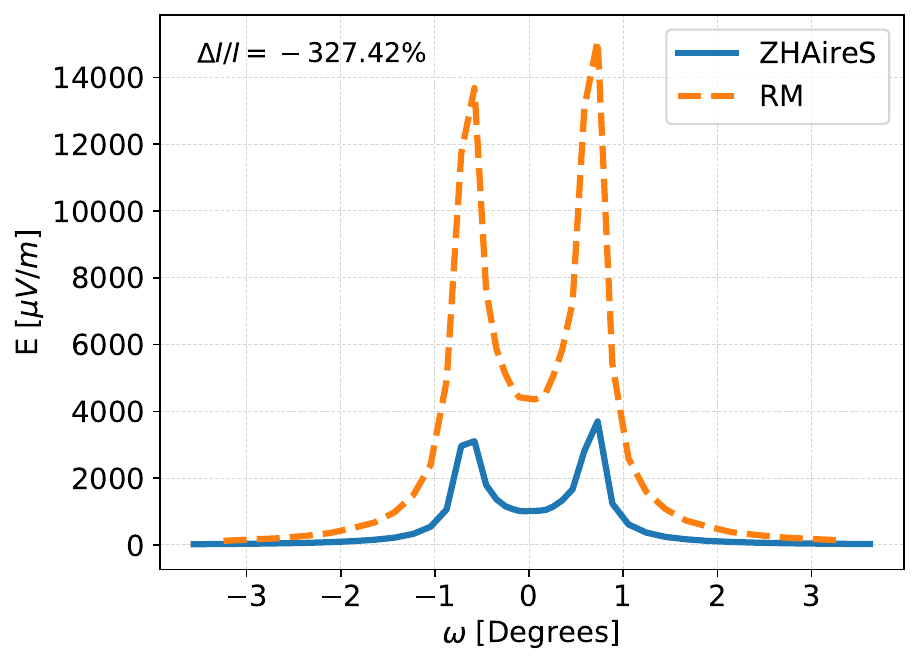}
    \end{subfigure}
    \hfill
    \begin{subfigure}[t]{0.45\textwidth}
        \includegraphics[width=\linewidth]{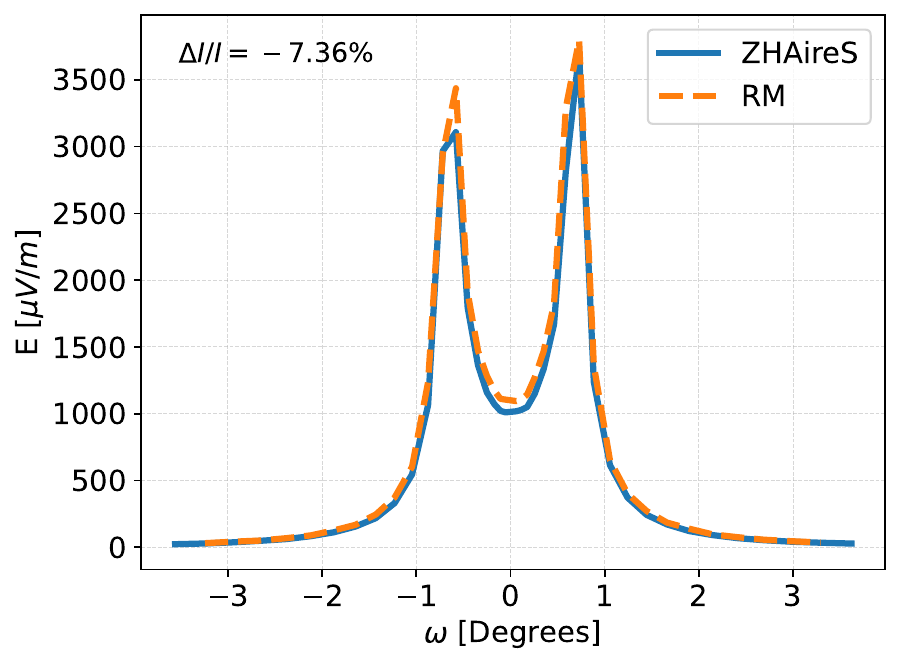}
    \end{subfigure}

    \vspace{0.5cm}

    \begin{subfigure}[t]{0.45\textwidth}
        \includegraphics[width=\linewidth]{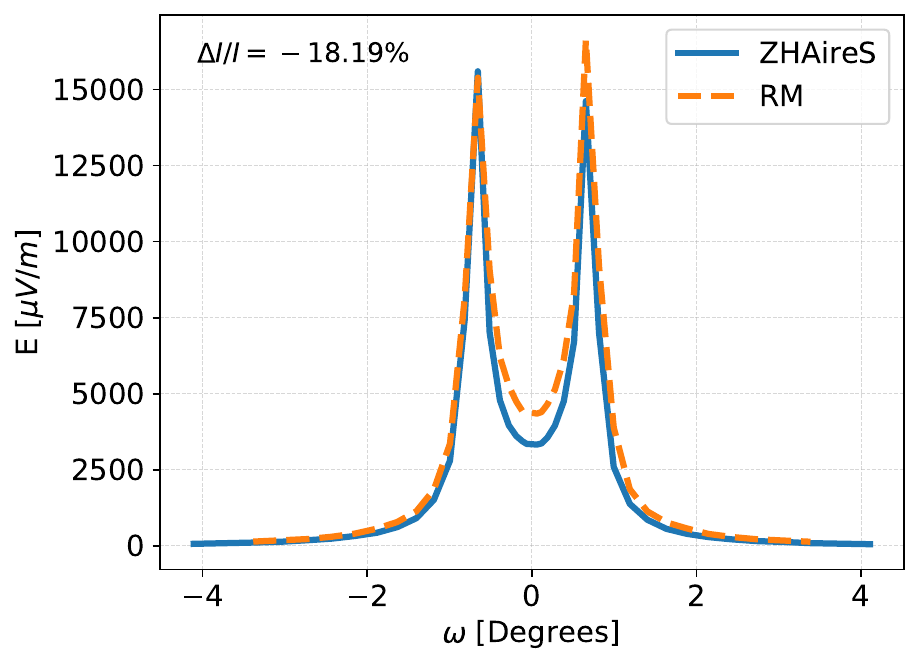}
    \end{subfigure}
    \hfill
    \begin{subfigure}[t]{0.45\textwidth}
        \includegraphics[width=\linewidth]{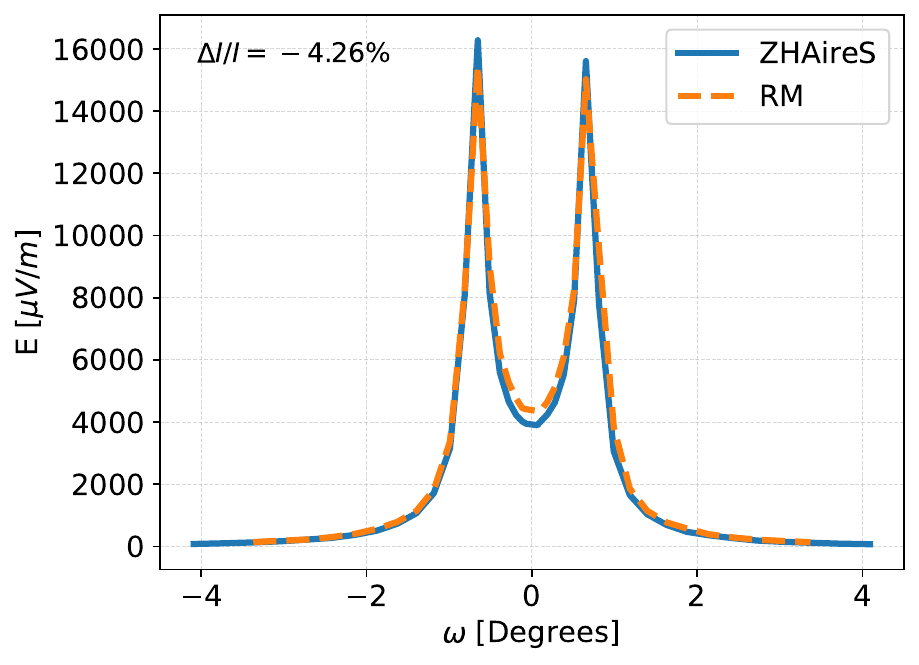}
    \end{subfigure}

    \caption{({\it Top}) Lateral distribution functions (LDFs) for two showers with the same arrival directions ($\theta= 75^{\circ},\, \varphi=90^{\circ}$), but different primary energies ($E_r = 3.98\, \rm EeV,\, E_t = 1.0\, \rm EeV $) without any energy scaling ({\it top left}) and with the energy scaling procedure ({\it top right}). The $\omega$ angle corresponds to the angular deviation from the shower axis, taken from $X_{\rm max}$ as shown in Fig.~\ref{fig:sketch_rm}.  $\Delta I/I$ gives the relative differences between the integral of both curves (see text). ({\it Bottom}) LDFs for two showers with the same primary energy and zenith angle ($E=3.98\, \rm EeV, \theta= 75^{\circ}$), but different azimuth angles ($\varphi_r = 90^{\circ},\, \varphi_t= 0^{\circ}$), with ({\it left}) and without ({\it right}) applying the scaling procedure.}
    \label{fig:E_phi_scaling}
\end{figure*}

with $E$ being  the shower primary energy. The left-hand panel of Fig.~\ref{fig:XmaxParam} shows the mean $X_{\rm max}$ values obtained from our simulation set together with the corresponding fits. We find $a_p =57.4 \, \rm g.cm^{-2}$, $c_p =421.9 \, \rm g.cm^{-2}$ for proton-induced showers with a coefficient of correlation $R_p^{2}=0.997$ and $a_{\rm fe} = 65.2 \, \rm g.cm^{-2}$, $c_{\rm fe} = 270.6 \, \rm g.cm^{-2}$ for iron showers, with a coefficient of correlation $R_{\rm fe}^{2}=0.999$. Using the same simulation set, we also parametrized the standard deviations of $X_{\rm max}$ following
\begin{equation}
    \sigma_{\rm xmax} = a + b/E^c \ .
\end{equation}
The right-hand panel of Fig.~\ref{fig:XmaxParam} shows the standard deviation of $X_{\rm max}$ and corresponding fits which  yield  $a_p =66.5  \, \rm g.cm^{-2}$, $b_p =2.84  \, \rm g.cm^{-2}.(E[TeV])^{c}$, $c_p =0.48$ for proton primaries with a coefficient of corelation $R_p^2=0.82$, and $a_{\rm fe} =20.9  \, \rm g.cm^{-2}$, $b_{\rm fe} =3.67 \, \rm g.cm^{-2}.(E[TeV])^{c_{\rm fe}}$, $c =0.21$ for iron primaries  with a coefficient of corelation $R_{\rm fe}^2=0.855$. These paramaetrizations of $ \sigma_{\rm xmax}$ are then used to model shower-to-shower fluctuations, as further discussed in Section~\ref{sec:fluctuations}. We note that these parametrizations are easily tunable within the Radio Morphing source code, which could be done upon demand.

Given a generated $X_{\rm max}$ grammage, we can then derive the $X_{\rm max}$ coordinates of the target shower by numerically integrating the grammage along the shower axis, assuming Linsley's atmospheric model. This $X_{\rm max}$ position is then assumed to be the point-like source of the target shower radio emission.  We mention that we used  here a parametrization of $<X_{\rm max}>$ based on numerical simulations, instead of an empirical one, in order to get results as close as possible to ZHAireS simulations. The current Radio Morphing implements only proton and iron primaries, but it would be straightforward to implement other cosmic ray primaries using a similar parametrization.

\subsection{Energy scaling}

 In air showers, the number of particles at $X_{\rm max}$ is expected to increase linearly with the primary particle energy (\cite{Engel2011doi:10.1146/annurev.nucl.012809.104544}). Additionally, for a coherent emission, the radiation energy scales with the squared number of emitters. This implies that at first order, the radio signal electric field amplitude at the antenna level increases linearly with the primary particle energy. Following~\cite{Zilles2020APh...114...10Z}, the scaling of the electric field amplitude $\mathcal{E}$ between a reference shower with primary energy $E_r$ and the same target shower but with primary energy $E_t$ is given by 
\begin{equation}
    \mathcal{E}^t(t, \vec{x}, \theta, \varphi) =  \frac{E_t}{E_r}\mathcal{E}^r(t, \vec{x}, \theta, \varphi) =  k_E \mathcal{E}^r(t, \vec{x}, \theta, \varphi) \ , \label{eq:Escaling}
\end{equation}
where we introduced $k_E =E_t/E_r$, and where $\vec{x}$ refer to the antenna position. We note here that rigorously, we should scale the electric field traces with the shower electromagnetic energy rather than with the primary energy, as hadrons  for example, do not contribute to the radio emission. Yet, the correction is minor, below $10\%$, and we assumed that it can be neglected with respect to the other sources of uncertainties in the Radio Morphing method.

We illustrate the scaling procedure in the top plots of Fig.~\ref{fig:E_phi_scaling}, where we show the lateral distribution function (LDF) of the radio signal along the $(\mathbf{v} \times \mathbf{B})$ direction, for two showers with different energies but the same directions, before (left-hand panel) and after (right-hand panel) the scaling. It can be seen that the LDF shows a much better agreement after the scaling procedure is applied. We can quantify the agreement between both showers by computing the relative differences in the integral of their LDFs noted $I$. In the case where no scaling is applied, we find $\delta I/I = (I_{\rm ZHS} - I_{\rm RM})/I_{\rm ZHS} = 327\%$, while $\delta I/I = -7\%$ after the scaling procedure.

\begin{figure*}[tb]
\includegraphics[width=0.49\linewidth]{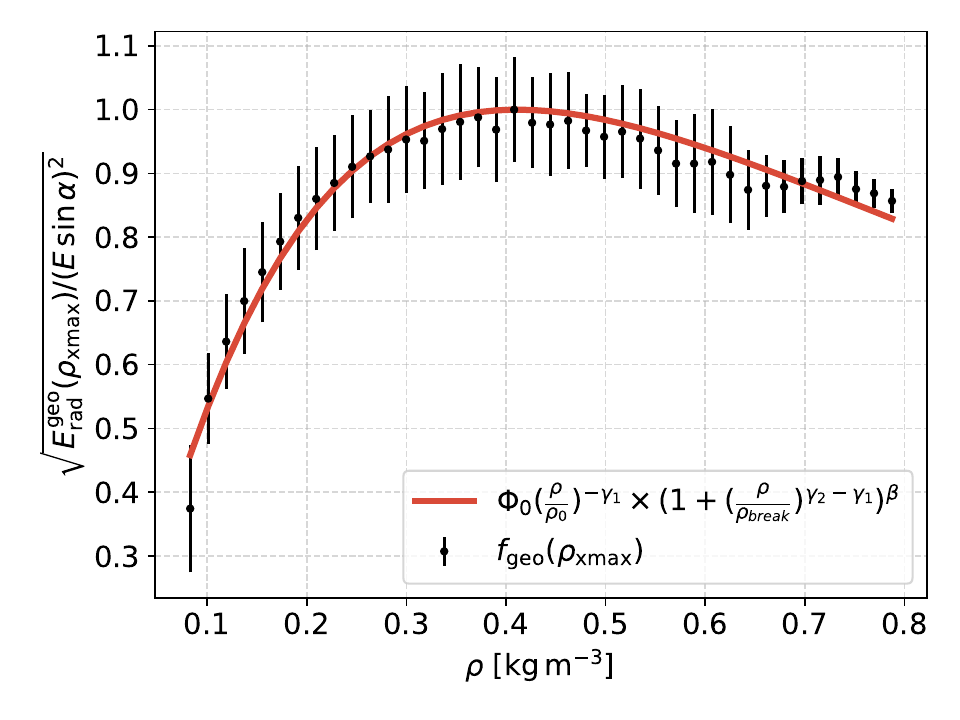}\hfill
\includegraphics[width=0.49\linewidth]{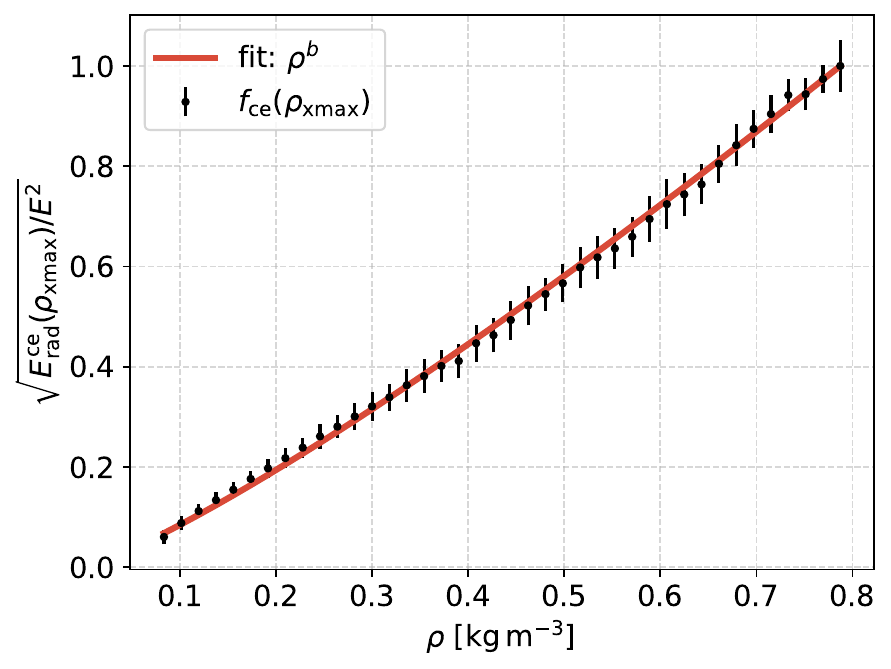}
\caption{({\it Left}) Average geomagnetic  electric field amplitude dependency with air-density (black crosses), from ZHAireS simulations. Results are fitted with a broken power law (red line), using $\phi_{0} = 1010$, $\gamma_{1} = -1.0047$, $\gamma_{2} = 0.2222$, $\rho_{\rm break} = 3.5\times 10^{-4}$, $\beta = 2.5$  and $\rho_{0} = 1.87\times 10^{-6}$. ({\it Right}) Average charge excess electric field amplitude dependency with air-density from ZHAireS simulations. Results are fitted with a simple power law, using $b = 1.195$. Both plots are normalized by their maximal value.}\label{fig:zenith_scaling}
\end{figure*}

\subsection{Geomagnetic angle}

A change in the shower direction will have two main effects: a change in the geomagnetic angle  $\alpha = \arccos(\mathbf{u_v}\cdot \mathbf{u_B})$, with $\mathbf{u_v}$ and $\mathbf{u_B}$ the shower and magnetic field direction respectively, and a change in the altitude at $X_{\rm max}$. The scaling with the geomagnetic angle incorporates all the effects related to the change in the shower azimuth, while the scaling with the altitude, discussed in the next section, is purely related to a primary composition and zenith angle effect.

Two main processes account for the radio emission  in air showers, the dominant geomagnetic and  the subdominant charge-excess mechanisms. Unlike the charge-excess component, which stays constant with the azimuthal direction, the geomagnetic emission varies with the shower direction since it is proportional to the Lorentz force and scales as $\mathcal{E}_{\rm geo} \propto qvB\sin{\alpha}$, with $\alpha$ the geomagnetic angle. Hence, to account for the variation of the radio emission with the shower azimuth, we have to separate the geomagnetic and charge-excess contributions at the antenna level and scale the geomagnetic component only. 
In practice, such a separation can be done using symmetries of the radio signal in the shower plane $[\mathbf{v} \times \mathbf{B}; \mathbf{v} \times (\mathbf{v} \times \mathbf{B})]$, which yields~\cite{HuegeEarlyLate2019ICRC...36..294H,AugerPolar2016PhRvL.116x1101A}
\begin{eqnarray}
    \mathcal{E}_{\rm ce} &=& \frac{\mathcal{E}_{ v \times (v \times B) }}{|\sin{\phi_{\rm obs}}|} \ , \label{eq:norm_ce}\\ 
    \mathcal{E}_{\rm  geo} &=& \mathcal{E}_{ v \times B}  - \mathcal{E}_{ v \times (v \times B)}\frac{\cos{\phi_{\rm obs}}}{|\sin{\phi_{\rm obs}}|} \ , \label{eq:norm_geo}
\end{eqnarray}
with $\phi_{\rm obs}$ is the angle between the antenna position in the shower plane and the $(\mathbf{v}\times \mathbf{B})$ axis. This equation diverges along the $v \times B$ axis $(\phi_{\rm obs} =0)$, but it can be generalized to any antenna position by using the specific case where $\phi_{\rm obs} =\pi/2$, corresponding  to antennas along the $v \times (v \times B)$ axis, and assuming  a radial symmetry of the radio signal in the shower plane. The assumption of radial symmetry remains justified, even for inclined air showers propagating through a density gradient, as studies suggest that radio signals propagate nearly along straight lines even for showers with zenith angles up to $\theta=88^{\circ}$, and frequencies below $600 \, \rm MHz$~\cite{Dieder2025PhRvD.111f3062V}.

The scaling of the radio emission with the geomagnetic angle affects only the $(v \times B)$ component of the traces and is then given by 
\begin{equation}
    \mathcal{E}_{v \times B}^{t}(t, \omega) =  -[k_{\rm geo}\,\mathcal{E}_{\rm geo}^{r}(t, \omega) + \cos{\phi_{\rm obs}}\,\mathcal{E}_{\rm ce}^{r}(t, \omega)] \ ,
\end{equation}
where $k_{\rm geo} = \sin{\alpha_{t}}/\sin{\alpha_{r}}$, $\mathcal{E}_{\rm geo}^{r}(t, \omega)$ and $\mathcal{E}_{\rm ce}^{r}(t, \omega)$ are the electric field amplitudes related to the geomagnetic and charge-excess emissions for the reference shower, evaluated at an angle $\omega$ from the shower axis and  for an observer angle $\phi_{\rm obs}$.

For a typical magnetic field $B=56\, \rm \mu T$, the charge-excess emission accounts for only $\sim 5\%$ of the signal amplitude for showers with zenith angle $\theta = 60^{\circ}$, and decreases with increasing zenith angle, becoming almost negligible for showers with zenith angle $\theta >80^{\circ}$~\cite{Chiche2022APh...13902696C}. Hence, for the most inclined showers, assuming a radial symmetry to separate the geomagnetic and charge-excess emissions may rather deteriorate the Radio Morphing accuracy. Consequently, for showers with $\theta > 70^{\circ}$ or in the case where the scaling correction factor is small $|1-k_{\rm geo}|<0.2$ we do not disentangle between the geomagnetic and charge-excess emission and scale directly the $(v \times B)$ component of the electric field following
\begin{equation}
    \mathcal{E}_{v \times B}^{t}(t, \omega) =  k_{\rm geo}\, \mathcal{E}_{v \times B}^{r}(t, \omega) \ .
\end{equation}

Similarly to what was done with the energy scaling, we show in the bottom plots of Fig.~\ref{fig:E_phi_scaling} the $(v \times B)$ component of the LDF along the $[\mathbf{ v \times (v\times B)}]$ direction, for a reference and a target shower with a different azimuth angle, before and after the scaling procedure. Here again, we can estimate the efficiency of the method by computing the relative differences in the integral of the two LDF. Much better agreement is found between the two showers, once the scaling procedure has been applied, with $\delta I/I=-18.19\%$ before the scaling and $\delta I/I = -4.3\%$ after scaling. The modeling of the charge-excess emission is a major improvement compared to~\cite{Zilles2020APh...114...10Z} which was assuming a pure geomagnetic emission.

\subsection{Altitude at \texorpdfstring{$X_{\mathrm{max}}$}{Xmax}}~\label{sec:xmax_scaling}

As mentioned in Section~\ref{sec:primary_scaling}, a change in the primary composition or in the shower zenith angle will change the altitude at $X_{\rm max}$ since inclined air showers are expected to develop higher in the atmosphere than vertical ones. Such a variation will impact both the refractive index and the air density at $X_{\rm max}$.

The variation of the air refractive index $n_{\rm air}$ at $X_{\rm max}$ was already modeled in~\cite{Zilles2020APh...114...10Z}. The refractive index is directly linked to the Cherenkov angle through $\theta_c = \arccos{1/n_{\rm air} \, \beta}$, which yields the characteristic scale of the radio signal amplitude pattern. We can account for modifications of $n_{\rm air}$ by stretching the reference antenna positions by a factor given by the ratio of the Cherenkov angle between the target and the reference shower. We get $\mathbf{x^{t}} =k_{c}\mathbf{x^{r}}$ with $k_{c} = \theta_{c}^{t}/\theta_{c}^{r}$. Since the shower energy must be conserved, we also scale the electric field traces with the inverse factor $\mathcal{E}^{t}(\mathbf{x^t}, t) = \frac{\mathcal{E}^{r}(\mathbf{x^r}, r)}{k_c}$.

Variations in the shower altitude also affect the air density at $X_{\rm max}$. The first version of Radio Morphing was limited to a narrow zenith-angle range, being valid only for nearly horizontal showers ($\theta \gtrsim 86^{\circ}$). In the updated version, we account for the dependence of the radio emission on the air density at $X_{\rm max}$, which extends the range of validity to zenith angles between $60^{\circ}$ and $87.1^{\circ}$.  Following~\cite{Chiche2022icrc.confE.194C}, we model this dependency using a set of $11\,000$ ZHAireS simulations of cosmic-ray showers (with proton or iron primaries) on a star-shape layout, with 13 zenith angles between $45^{\circ}$ and $89^{\circ}$, 17 primary particle energies between $\rm 0.01\,EeV$ and $\rm 3.98\,EeV$ and 3 azimuth angles ($0^{\circ}, 90^{\circ}, 180^{\circ}$). For each air shower, we reconstructed the charge excess and geomagnetic radiation energies $E_{\rm rad}^{\rm ce}$ and $E_{\rm rad}^{\rm geo}$, by separating each component and integrating their energy fluence following~\cite{Glaser2016JCAP...09..024G, Chiche2022icrc.confE.194C}. We note that the radiation energy is a calorimetric observable and therefore does not depend on the distance to $X_{\rm max}$. The radiation energies were then corrected from any dependency on the primary particle energy and on the geomagnetic angle using~\cite{Engel2011doi:10.1146/annurev.nucl.012809.104544,Kahn1966RSPSA.289..206K, Glaser2016JCAP...09..024G}
\begin{equation}
     \tilde{E}_{\rm rad}^{\rm geo} = \frac{E_{\rm rad}^{\rm geo}}{(\mathcal{E} \times \sin{\alpha})^{2}} \qquad {\rm and} \qquad \tilde{E}_{\rm rad}^{\rm ce} = \frac{E_{\rm rad}^{\rm ce}}{\mathcal{E}^{2}} \ ,
\end{equation}
with $\mathcal{E}$, the primary energy and where a correction in $\mathcal{E}^{2}$ is applied since the radiation energy scales with the square of the electric field, which itself is proportional to the primary energy.
 The residual variations of $\tilde{E}_{\rm rad}^{\rm geo}$ and $\tilde{E}_{\rm rad}^{\rm ce}$ should thus stem from density effects only. Finally, since the electric field scales with the square root of the radiation energy, the geomagnetic and charge-excess electric field dependency with air density at $X_{\rm max}$ can be derived from $\sqrt{\tilde{E}_{\rm rad}^{\rm geo}(\rho_{\rm xmax})}$ and $\sqrt{\tilde{E}_{\rm rad}^{\rm ce}(\rho_{\rm xmax})}$. Assuming the Linsley atmospheric model, we can fit these dependencies with
 \begin{eqnarray}
     f_{\rm geo}(\rho_{\rm xmax}) &=& \phi_{0}\left(\frac{\rho}{\rho_0}\right)^{-\gamma_1}\left[1+\left(\frac{\rho}{\rho_{\rm break}}\right)^{\gamma_2-\gamma_1}\right]^\beta \\
     f_{\rm ce}(\rho_{\rm xmax}) &=& \rho^{b}
 \end{eqnarray} \,
  using $\phi_{0} = 1010$, $\gamma_{1} = -1.0047$, $\gamma_{2} = 0.2222$, $\rho_{\rm break} = 3.5\times 10^{-4} {\, \rm kg.m^{-3}}$, $\beta = 2.5$  and $\rho_{0} = 1.87\times 10^{-6}{\, \rm kg.m^{-3}}$ for the geomagnetic component, which yields a coefficient of correlation $R^2_{\rm geo}=0.960$ and with $b=1.195$ for the charge-excess component, which gives a coefficient of correlation $R^{2}_{\rm ce}=0.998$.

In Fig.~\ref{fig:zenith_scaling}, we show the mean geomagnetic and charge-excess emission dependency with air density at $X_{\rm max}$, where the results for the full set of $11\,000$ simulations where averaged in evenly spaced air density bins. The right-hand panel shows that the charge-excess amplitude increases almost linearly with the air density, as also shown in~\cite{Glaser2016JCAP...09..024G}. Indeed, a higher density provides more air atoms that can be ionized by the shower, leading to a larger build-up of negative charges, resulting in a stronger charge-excess emission. The left-hand panel shows that the geomagnetic emission first increases and then decreases when decreasing the air density at $X_{\rm max}$, i.e., for more inclined showers. When the shower inclination increases, charged particles can reach a larger drift velocity which initially results  in a stronger transverse current and a larger geomagnetic emission~\cite{Scholten2008APh....29...94S}. Yet, if the shower is too inclined and the magnetic field  is strong enough, the shower lateral extent at $X_{\rm max}$ becomes so large that the radio emission becomes incoherent, causing the drop observed for the lowest air densities as characterized in~\cite{chiche2024PhRvL.132w1001C}. As a result, the charge-excess emission was fitted using a simple power law, while the geomagnetic emission was fitted using a broken power-law. These fits were done using the full band electric field, to get an average trend of the coherence loss effects.

\begin{figure}[t]  
    \centering
    \includegraphics[width=\linewidth]{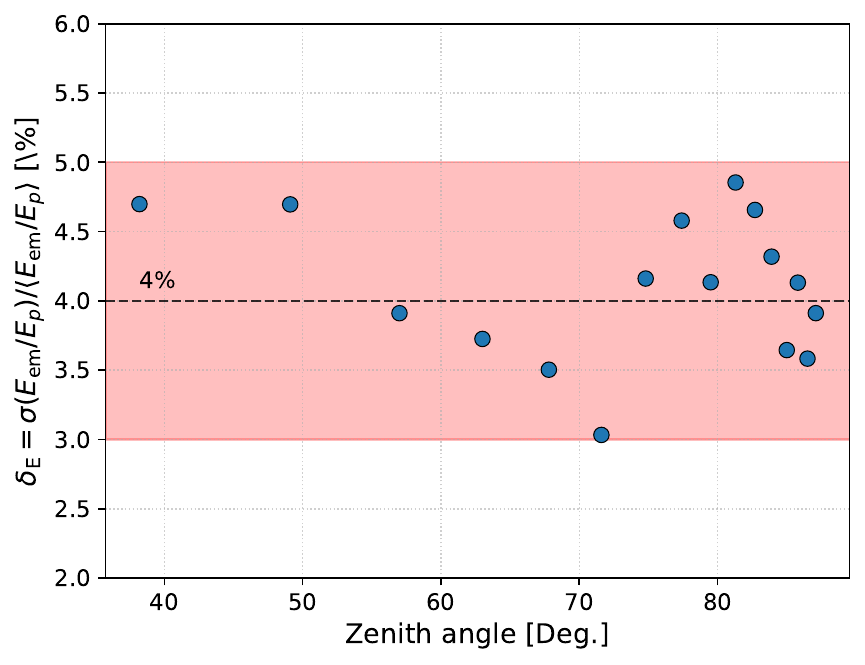}
    \caption{Fluctuation of the electromagnetic energy $E_{\rm em}$ normalized to the primary particle energy as a function of zenith angle, from ZHAireS simulations. The red band show that fluctuations are roughly constant with zenith angle and contained between $3\%$ and $5\%$, with an average around $4\%$ (black dashed line).} 
    \label{fig:fluctuations}
\end{figure}

\subsection{Shower-to-shower fluctuations}\label{sec:fluctuations}

 Due to the probabilistic nature of interactions occurring in air-showers, the radio-emission from various showers induced by primary particles with identical mass composition, energy, and arrival direction, presents some differences that we refer to as shower-to-shower fluctuations. These fluctuations are intrinsically modeled by Monte-Carlo simulations for which hadronic interactions are evaluated in a probabilistic way. 
However, Radio Morphing uses only a small sample of reference showers; hence, hadronic interactions are not recomputed when estimating radio signals. This allows one to considerably reduce the computation time but requires further implementation of shower-to-shower fluctuations that are consistent with the simulation data.  This new Radio Morphing version models these effects, at the first order, with parameterizations of $X_{\rm max}$ fluctuations and fluctuations in the number of particles in the shower. The user can now enable shower-to-shower fluctuations when generating radio signal estimations.


We used the parametrizations of $\sigma_{\rm xmax}$ shown in Section~\ref{sec:primary_scaling} to model the dispersion of $X_{\rm max}$ around its average value. When shower-to-shower fluctuations are enabled by the user, the $X_{\rm max}$ grammage of the target shower used by Radio Morphing is drawn from a Gaussian distribution, using $\langle X_{\rm max} \rangle$ and $\sigma_{\rm xmax}$ as input parameters.

We use the parametrizations of $\sigma_{\rm xmax}$ presented in Section~\ref{sec:primary_scaling} to model the dispersion of $X_{\rm max}$ around its mean value. When shower-to-shower fluctuations are enabled by the user, the $X_{\rm max}$ grammage of the target shower used in Radio Morphing is randomly drawn from a Gaussian distribution using $\langle X_{\rm max} \rangle$ and $\sigma_{\rm xmax}$ as input parameters.



We also studied the fluctuation of the electromagnetic energy in air-showers using ZHAireS simulations. For each simulation, we computed the electromagnetic energy $E_{\rm em}$ at $X_{\rm max}$ that we normalized by the primary energy $E_{\rm p}$. Fluctuations of the electric field amplitude are then evaluated from the ratio between the standard deviation and the average value of the distributions
\begin{equation}
    \delta_E(\theta)= \frac{\sigma(E_{\rm em}/E_p)}{\langle E_{\rm em}/E_p \rangle } \ .
\end{equation}
In Fig.~\ref{fig:fluctuations}, we show the results on the fluctuations of the electromagnetic energy as a function of the shower zenith angle. We find fluctuations around 4\% independently of the zenith angle. In Radio Morphing, we can model the fluctuations of $E_{\rm em}$ by introducing fluctuations of the target shower primary particle energy, since both quantities scale linearly. In practice, this is achieved by randomly drawing a shower energy from a Gaussian distribution centered on the target energy $E_t^{\rm input}$, specified by the user, with a standard deviation $\delta_E =4\%$.  When shower-to-shower fluctuations are enabled, the effective target energy therefore follows
\begin{equation}
    E_t^{\rm eff} = E_t^{\rm input} + \delta_E E_t^{\rm input}G \ ,
\end{equation}
where  $G=\mathcal{N}(0,1)$ is a standard normal random variable.

%


\begin{figure*}[tb]
\includegraphics[width=0.49\linewidth]{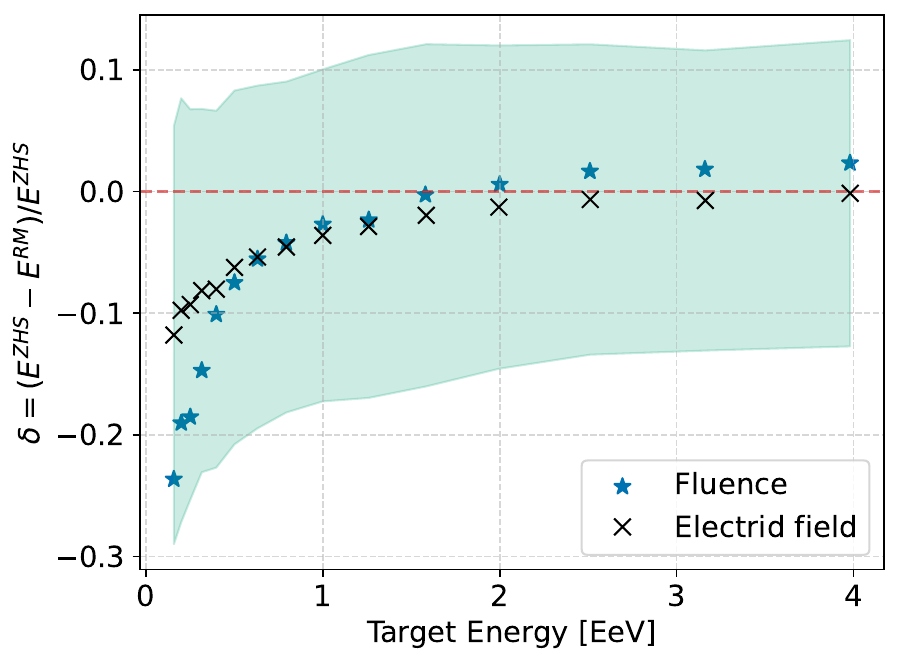}\hfill
\includegraphics[width=0.49\linewidth]{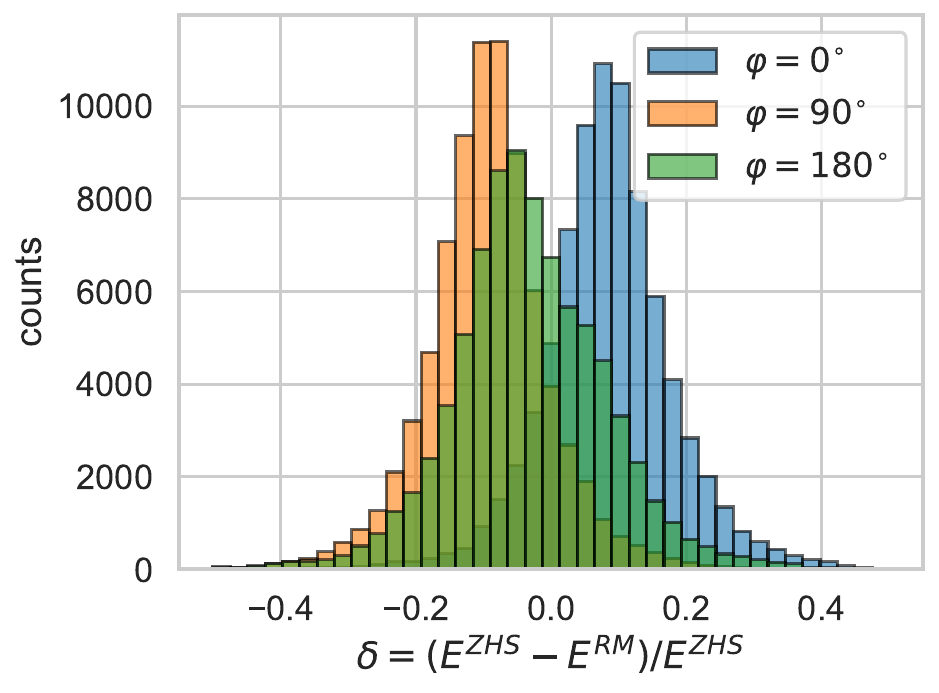}
\caption{({\it Left}) Relative differences $\delta$ on the electric field peak amplitude (black crosses) and on the energy fluence $\delta_{\rm fluence} = (f^{\rm ZHS}-f^{\rm RM})/f^{\rm ZHS}$ (blue stars) between ZHAireS and Radio Morphing simulations as a function of the shower energy. Crosses show the mean value of $\delta$ while the shaded area corresponds to its standard deviation. ({\it Right}) Distribution of $\delta$ at the antenna level for showers with three different azimuth angles.}\label{fig:RMperf_vs_E}
\end{figure*}

\begin{figure*}[tb]
\includegraphics[width=0.49\linewidth]{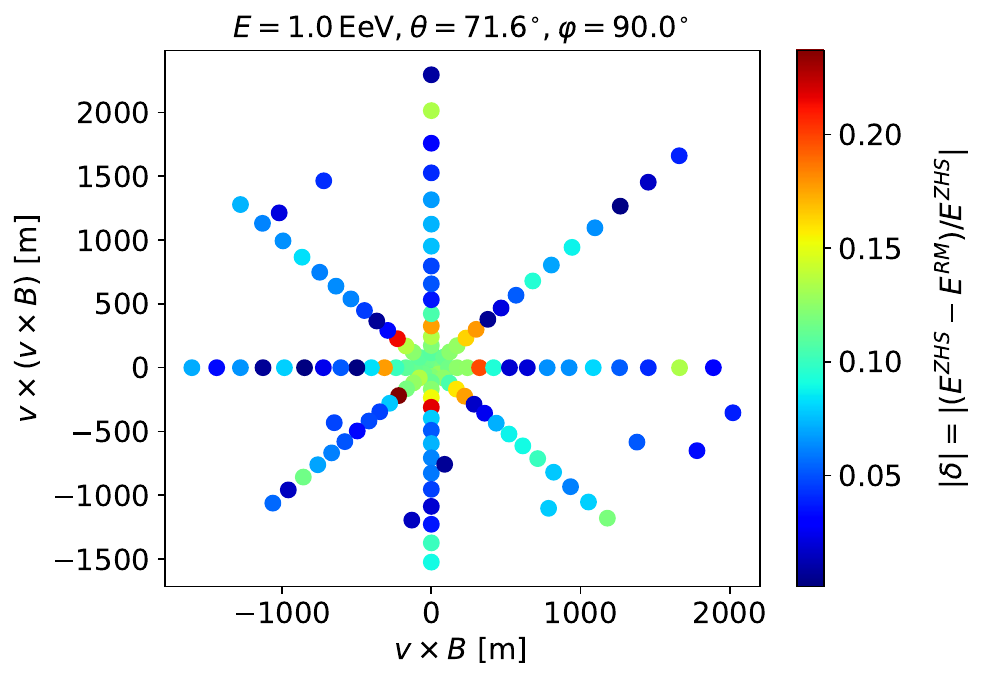}\hfill
\includegraphics[width=0.49\linewidth]{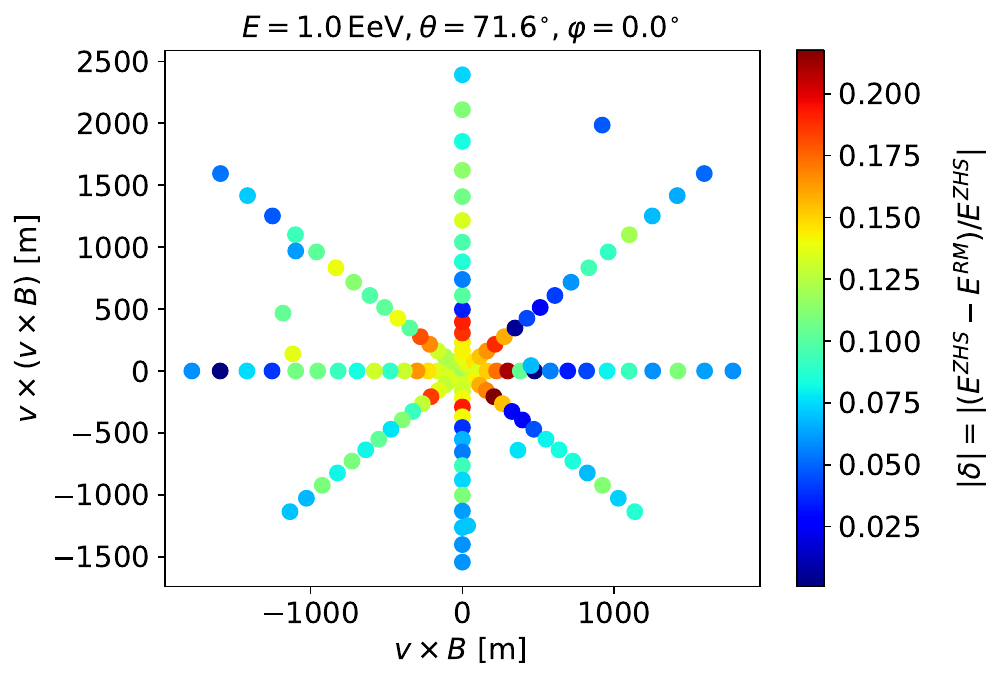}
\caption{2D maps showing the spatial distribution of the relative error $|\delta|= |E^{\rm ZHS}-E^{\rm RM}|/E^{\rm ZHS}$ (color scale) between ZHAireS and Radio Morphing simulations on a star-shaped antenna grid,  for two representative showers  with $\varphi=90^\circ$ ({\it left}) and $\varphi=0^\circ$ ({\it right}) .}\label{fig:RMphi_delta_distrib}
\end{figure*}

\subsection{Interpolation and extrapolation}

Once the scaling procedure is completed, we perform a 2D interpolation of the radio signal in the scaled plane of antennas illustrated in Fig.~\ref{fig:sketch_rm}, to infer the radio emission at any desired position within the plane. For this purpose, we rely on the method presented in~\cite{Tueros2021JInst..16P2031T} where the traces are decomposed into a phase and amplitude in Fourier space which are then interpolated independently using the closest neighbor antennas.  When considering antennas away from the Cherenkov angle, the method provides relative differences with ZHAireS simulations of only a few percent on both the peak-to-peak amplitude and  integrated pulse, combined with a timing error of less than 5 nanoseconds. This is a major improvement compared to the initial version of the Radio Morphing implementation (\cite{Zilles2020APh...114...10Z}).

Eventually, to evaluate the radio emission at various longitudinal distances, the 2D interpolation is coupled with a 3D extrapolation, as illustrated in Fig.~\ref{fig:sketch_rm}. In practice, for any target antenna at a distance $d_{\rm ant}^{\rm Xmax}$ from $X_{\rm max}$, with an angular deviation from the shower axis $\omega$  and an observer angle $\phi_{\rm obs}$, we first perform a 2D interpolation within the scaled plane for the antenna with the same ($\omega$, $\phi_{\rm obs}$) angles. The radio emission at the target antenna is then inferred from the scaled one by performing an extrapolation that accounts for propagation effects (propagation of the blue dashed line from the red scaled plane in Fig.~\ref{fig:sketch_rm} to the extrapolated signal on ground). This modeling assumes that the radio emission amplitude pattern in the shower plane, in the ($\omega$, $\phi_{\rm obs}$) representation, is  invariant by translation along the shower axis.

The correction for the propagation effects assumes a point-like emission at $X_{\rm max}$, so that the radio emission electric field amplitude decreases linearly with the distance to $X_{\rm max}$. For fixed ($\omega$, $\phi_{\rm obs}$) angles, considering  a scaled antenna at a distance $d_{\rm scaled}$ from $X_{\rm max}$ and a target antenna at a distance $d_{\rm target}$, our correction yields

\begin{equation}
    \mathcal{E}_{\rm target}(\omega, \, \phi_{\rm obs}) = \frac{d_{\rm scaled}}{d_{\rm target}}\mathcal{E}_{\rm scaled}(\omega, \, \phi_{\rm obs}) \ .
\end{equation}
A second correction consists in accounting for the difference in the integrated refractive index between the scaled and the extrapolated signals. Indeed, antennas at different longitudinal distances will see a different integrated refractive index from $X_{\rm max}$ to their position. This effect is corrected  by applying a stretching factor to the antenna positions and the traces $k_{\rm stretch} = \theta_{c}^{\rm int}/\theta_{c}^{\rm scaled} = (\arccos{1/\bar{n}_{\rm int}})/(\arccos{1/\bar{n}_{\rm scaled}})$, similarly to what was done for the modeling of the air density effects in the scaling with the zenith angle (Section~\ref{sec:xmax_scaling}).

\section{Radio Morphing Performances}~\label{sec:performances}

The Radio Morphing performances on the radio signal amplitude, timing, and frequency spectrum are evaluated by comparing Radio Morphing outputs to ZHAireS simulations run with the same input parameters (mass composition, primary energy, arrival direction, and antenna positions).  For this purpose, we generated with Radio Morphing $\sim 4500$ proton or iron showers in the energy range  $E= [0.2-3.98]\, \rm EeV$,  zenith angles between $\theta = [63^{\circ}-87.1^{\circ}]$, and for 3 azimuth angles $\varphi = [0^{\circ}, 90^{\circ}, 180^{\circ}]$. The lower bound of the explored energy range follows directly from the zenith-angle range targeted by Radio Morphing, as showers with lower primary energies typically do not trigger a sufficient number of antennas in standard radio arrays. The upper bound is mainly set by the current extent of our ZHAireS test-simulation library and will be extended in future studies. Nevertheless, given the steeply falling cosmic-ray energy spectrum, the selected energy range is expected to encompass the majority of showers triggering radio detectors.

\subsection{Amplitude results}

 For each Radio Morphing test simulation, the radio signal was generated at 176 antenna positions located at an altitude of 1086 meters above sea level corresponding to the location of Dunhuang (China), and compared to the same outputs produced with ZHAireS simulations. We then quantified the efficiency of our method by evaluating the relative differences in the electric field amplitude between ZHAireS and Radio Morphing simulations using
\begin{equation}
    \delta = \left \langle \frac{E^{\rm ZHS}(t) - E^{\rm RM}(t)}{E^{\rm ZHS}(t) } \right\rangle_{> 0.2 E_{\rm zhs}^{\rm peak}}  \ ,
\end{equation} 
where the brackets denote an average over time samples with amplitudes exceeding 20\% of the ZHAireS peak value, considering a sampling rate of $0.5\, \rm ns$. This metric was preferred over a simple comparison of peak amplitudes, as it remains valid even for filtered signals, where the power is distributed over several samples. We also note that for each antenna, the dominant channel is used to evaluate the value of $\delta$. Using this metric, the mean value of $\delta$ over different signals quantifies the accuracy of our method (bias), while the standard deviation of $\delta$ evaluates the precision of Radio Morphing.

In the left-hand panel of Fig.~\ref{fig:RMperf_vs_E}, we show the mean (crosses) and standard deviation (blue shaded area) of the relative differences $\delta$ in the total peak amplitude (black stars) and on the energy fluence (blue stars) using $\delta_{\rm fluence} = (f^{\rm ZHS}-f^{\rm RM})/f^{\rm ZHS}$, at the antenna level, between the ZHAireS and Radio Morphing simulations, as a function of the target shower energy. We get mean relative differences close to $\delta = 0 \%$ and standard deviations of $\sim \delta =10\%$ for the most energetic showers, with energies above $1\, \rm EeV$, while both the precision and accuracy of Radio Morphing decrease for the lowest-energy showers. This loss of efficiency at low energy comes from two main effects. First, as mentioned in Section~\ref{Sec:Method}, our reference showers were run with a primary energy of $\mathcal{E}_r = 3.98\, \rm EeV$, so the lowest primary energies in Fig.~\ref{fig:RMperf_vs_E} are also the farthest from the reference one. Second, at low energy the cascade involves fewer particle interactions, resulting in reduced statistical averaging of fluctuations and therefore a larger intrinsic dispersion in the radio signal and in the relative differences $\delta$. The results on the energy fluence show a similar trend to the ones on the electric field amplitude. Main differences are observed for the lowest primary particle energies where the relative differences on the fluence becomes larger and can be significant. This trend is expected as the energy fluence is  proportional to the squared of the electric field, but the precision of Radio Morphing on the fluence could be improved simply by adding an additional reference shower at lower primary particle energy. 

The right-hand panel of Fig.~\ref{fig:RMperf_vs_E} shows the distribution of $\delta$ for the three azimuth angles tested in our simulations, restricting to showers with primary energy above $2\, \rm EeV$. We observe a slight azimuth-dependent bias, as the distributions are not exactly centered on $\delta =0$, the largest deviations being observed for $\varphi = 0^\circ$. This effect is still under investigation. We hypothesize that the scaling of the radio emission with zenith angle shown in Fig.~\ref{fig:zenith_scaling} is in fact azimuth-dependent. Showers with smaller geomagnetic angles — for instance $\varphi = 0^\circ$ at the Dunhuang site — experience reduced coherence losses and therefore yield stronger radio emission than showers with larger geomagnetic angles~\cite{chiche2024PhRvL.132w1001C, Ammerman2023JCAP...08..015A}. The current zenith-angle scaling implemented in Radio Morphing, which accounts for coherence loss effects, effectively represents an average over all azimuth angles. This limitation can be addressed by extending the reference library to include showers covering a range of azimuth angles, or by introducing an azimuth-dependent zenith-angle scaling. Such improvements will be implemented in future developments of the method. To further investigate the differences in $\delta$ observed for $\varphi=0^{\circ}$ in Fig.~\ref{fig:RMphi_delta_distrib}, we show the spatial distributions of $|\delta|$ in the shower plane for two representative simulations with $\varphi=90^{\circ}$ (left-hand panel) and $\varphi=0^{\circ}$ (right-hand panel), both at a zenith angle of $\theta=71.6^{\circ}$. While the overall distributions are similar, the relative errors exhibit a less rotationally symmetric pattern for $\varphi=0^{\circ}$ than for $\varphi=90^{\circ}$. Interestingly, this asymmetry is mainly observed when $\theta_{\rm target} \neq \theta_{\rm ref}$, i.e. when the radio emission is also scaled with zenith angle. This strengthen the assumption that the observed azimuthal dependence is associated with limitations of the current zenith-angle scaling, which does not explicitly account for azimuth-dependent effects such as the clover-leaf emission pattern, and coherence-loss effects discussed in~\cite{chiche2024PhRvL.132w1001C}.

\begin{figure}[t]  
    \centering
    \includegraphics[width=\linewidth]{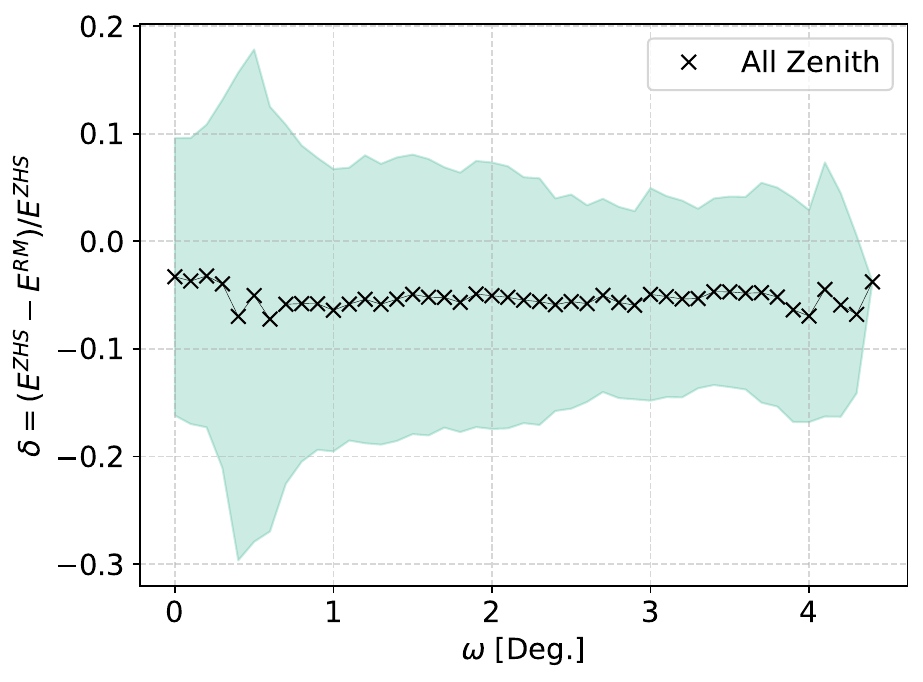}
    \caption{Mean (crosses) and standard deviation (shaded area) of the relative differences $\delta$ on the electric field peak amplitude, between ZHAireS and Radio Morphing simulations, as a function of the radio antenna positions. The $\omega$ angle corresponds to the angular deviation of the antennas from the shower axis.} 
    \label{fig:omega_scaling}
\end{figure}

In Fig.~\ref{fig:omega_scaling}, we show the mean (crosses) and standard deviation (shaded area) of $\delta$ as a function of the antenna angular deviation from the shower axis, $\omega$. The mean relative differences between ZHAireS and Radio Morphing remain roughly constant across antenna positions, which indicates that there is no significant bias related to the antenna positions. The standard deviations are also constant at large $\omega$, but a significant increase appears around $\omega = 0.6^{\circ}$, i.e., near the Cherenkov angle. This effect is mainly due to the reduced performance of the radio signal interpolation at the Cherenkov angle, where the spatial variations of the radio signal amplitude are stronger. This feature could be improved by using reference ZHAireS simulations with an even denser antenna grid close to the Cherenkov angle.

\begin{figure*}[tb]
\includegraphics[width=0.49\linewidth]{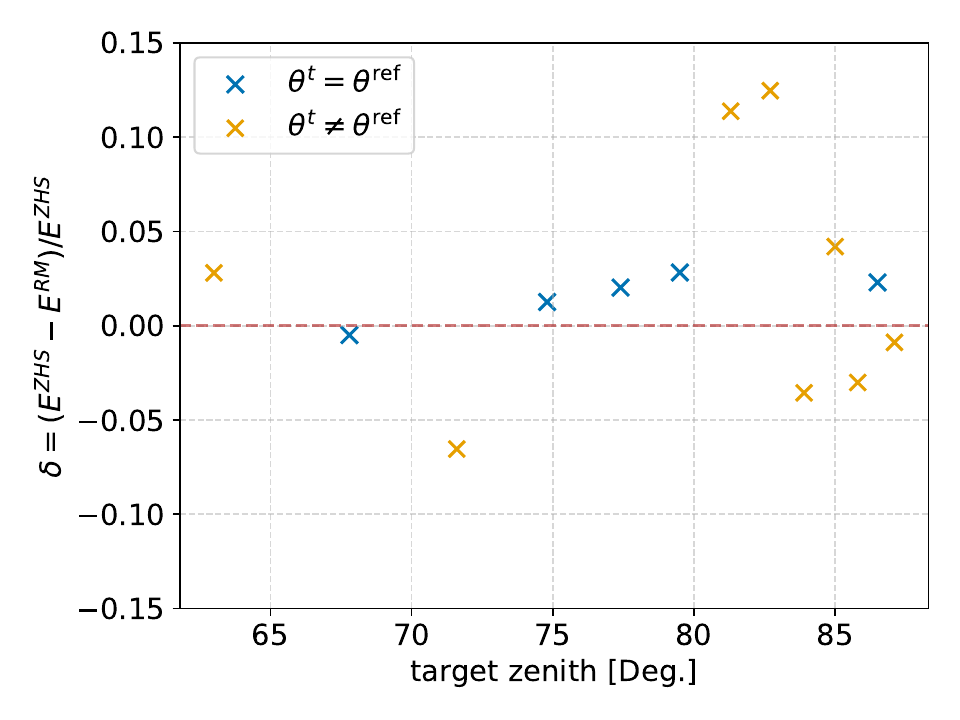}\hfill
\includegraphics[width=0.49\linewidth]{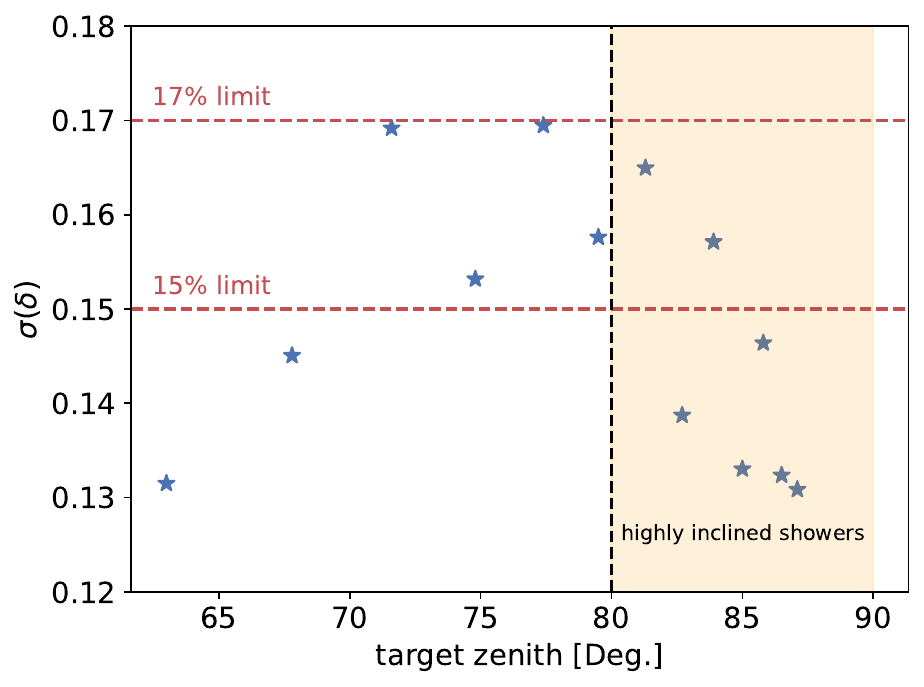}
\caption{Mean (left-hand panel) and standard deviation (right-hand panel) of the relative differences $\delta$ on the electric field peak amplitude, between ZHAireS and Radio Morphing simulations, as a function of the shower zenith angle, without any trigger threshold.}\label{fig:RMperf_vs_theta}
\end{figure*}

\begin{figure}[t]  %
    \centering
    \includegraphics[width=\linewidth]{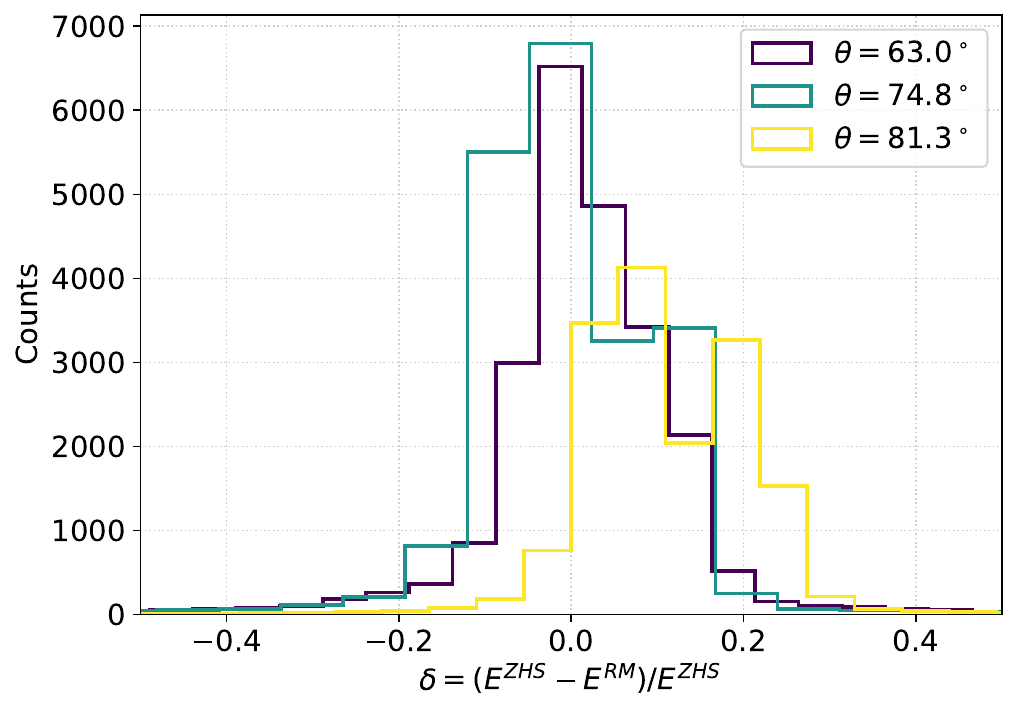}
    \caption{Distribution of the relative differences $\delta$ for a few representative zenith angles for unfiltered traces with a trigger threshold on the electric field amplitude of $110\, \rm \mu V/m$. } 
    \label{fig:delta_zenith_distrib}
\end{figure}

\begin{figure*}[!ht]
\includegraphics[width=0.49\linewidth]{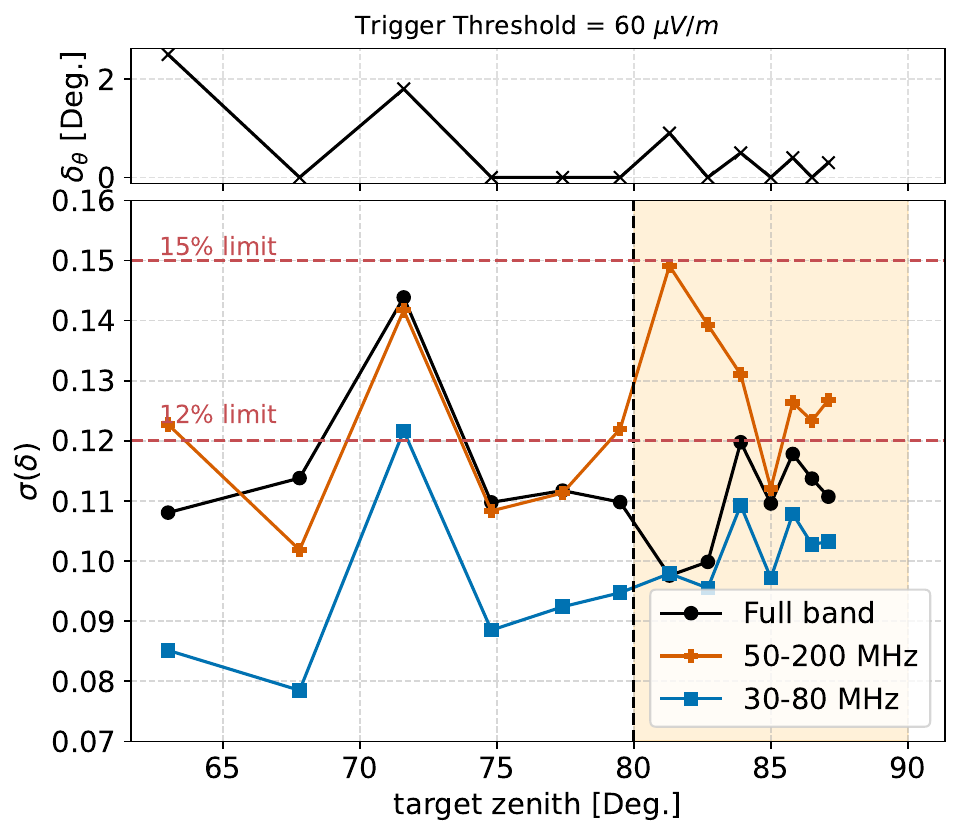}\hfill
\includegraphics[width=0.49\linewidth]{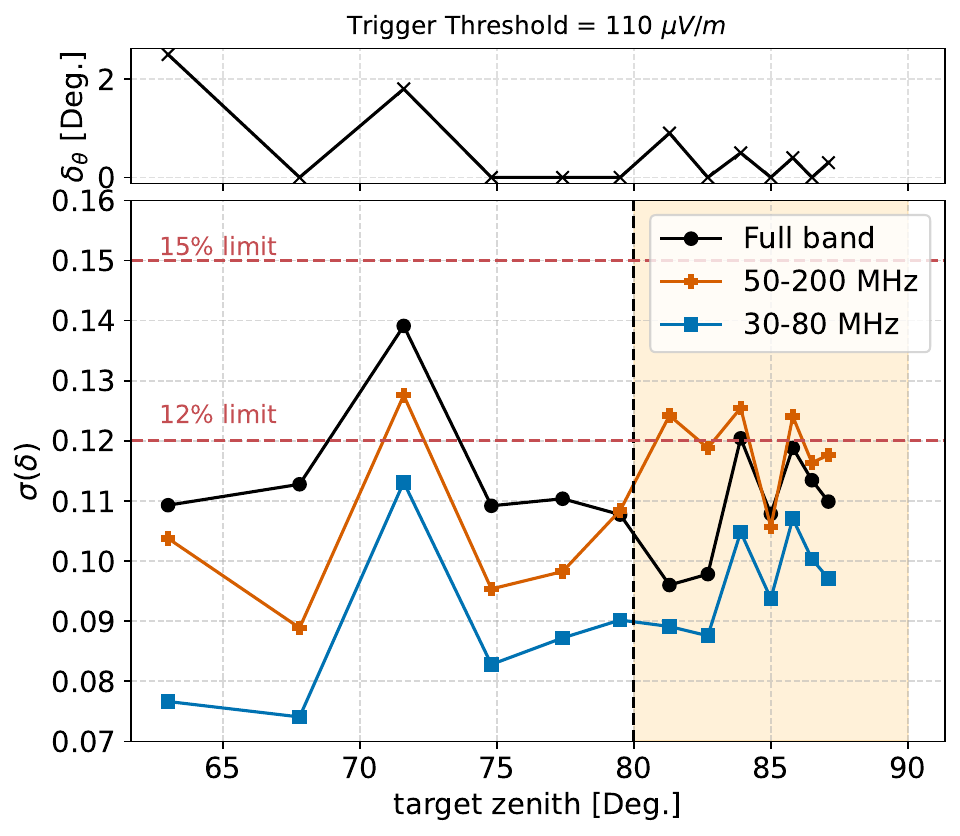}
\caption{Standard deviations of of the relative differences $\delta$ on the electric field peak amplitude, between ZHAireS and Radio Morphing simulations, as a function of the shower zenith angle. The results are shown for three different frequency bands: full band (black line), $[50-200]\, \rm MHz$ (orange line) and  $[30-80]\, \rm MHz$ (blue line). The left-hand panel corresponds to a $3 \sigma$ trigger threshold ($60 \, \rm \mu V/m$), while the right-hand plot is for a $5 \sigma$ trigger threshold ($110 \, \rm \mu V/m$). The shaded area delimits the most inclined showers ($\theta>80^{\circ})$. The top panels show the differences in zenith angle $\delta \theta$ between the reference shower used and the target shower for which the radio signal is generated.}\label{fig:RMstd_vs_theta}
\end{figure*}

In Fig.~\ref{fig:RMperf_vs_theta} we show the mean (left-hand panel) and standard deviations (right-hand panel) of $\delta$ as a function of the target shower zenith angle. For the average values of $\delta$ we distinguish between the case where the target zenith angle is the same as that of our reference showers (blue crosses) and the cases where both quantities are different (orange crosses).  For most zenith angles, Radio Morphing provides mean relative differences in the peak amplitude of less than $5\%$, evidencing a minor bias in our simulations. Larger deviations are observed only for showers with target zenith angles that are relatively far from any reference zenith angle in our library. Typically for $\theta = 72^{\circ}$, while the closest showers have either $\theta^{\rm ref} = 69.8^{\rm \circ}$ or $\theta^{\rm ref} = 74.8^{\rm \circ}$ . This highlights that the efficiency of Radio Morphing depends on the proximity of the target showers to the reference ones, but it also indicates that its performance could be improved by adding more reference showers. In Fig.~\ref{fig:delta_zenith_distrib} we also  show the distribution of the relative differences  $\delta$ for a few representative zenith angle values, after applying a trigger threshold of $110\, \rm \mu V/m$. We find that the distributions at $\theta=63^{\circ}$ and $\theta=74.8^{\circ}$ are peaked around $\delta=0$ while a global shift is observed for $\theta=81.3^{\circ}$. This confirms the observed bias in the left-hand panel of Fig.~\ref{fig:RMperf_vs_theta} and suggests that our scaling of the radio emission may sometimes systematically underestimates or overestimate ZHAireS radio signals for specific zenith angle values. The origin of this bias will be further investigated in future versions of Radio Morphing.

The right-hand panel of Fig.~\ref{fig:RMperf_vs_theta} shows that Radio Morphing achieves standard deviations of the peak-amplitude error below 17\% over the full zenith-angle range, and even below 12\% for the most vertical and the most inclined showers. This behavior can be understood as follows: vertical showers are generally easier to model, since their radio emission varies less steeply with zenith angle and their footprints are more symmetric. At large zenith angles, the improved precision originates from the higher density of reference showers in our library, which enhances the interpolation accuracy of the Radio Morphing method.

\subsection{Filtered results}

To further assess the performances of Radio Morphing, we evaluated its precision after applying frequency filtering and different trigger thresholds. Figure~\ref{fig:RMstd_vs_theta} shows the standard deviation of $\delta$ as a function of zenith angle for unfiltered traces, traces filtered in the $[50-200]\, \rm MHz$ band, and traces in the $[30-80]\, \rm MHz$  band. Assuming that the typical stationary background noise has a standard deviation of $22\, \rm \mu V/m$~\cite{Decoene2021NIMPA.98664803D}, a trigger threshold of $60 \, \rm \mu V/m$ ($3 \sigma$) was applied for the results shown in the left-hand panel, and $110\, \rm \mu V/m$ ($5 \sigma$) for the right-hand panel. To ensure a consistent comparison, all trigger thresholds were defined with respect to the signal filtered in the $[30-80]\, \rm MHz$ band. We find standard deviations below $15\%$ independently of the frequency band, trigger threshold, and zenith angle. A comparison with the top panel, which shows the difference in zenith angle between the reference and target showers, evidences that the largest fluctuations are observed for target zenith angles relatively far from those of the reference showers (typically for $\theta = 72^{\circ}$ and $\theta = 81.3^{\circ}$). This behavior is expected, since larger differences between the reference and target showers require Radio Morphing to rely more heavily on the scaling procedure, which is not perfectly accurate. We nevertheless emphasize that the accuracy of Radio Morphing is easily tunable and that, while the data points at $\theta = 72^{\circ}$ and $\theta = 81.3^{\circ}$ were kept for illustrative purposes, these fluctuations could be reduced simply by adding more showers close to these zenith angles to the reference library.

We further notice that the performances in the GRAND frequency band ($[50-200]\, \rm MHz$) are not necessarily better than the full band results. This effect seems to be linked mainly to low-amplitude signals, as for a trigger threshold of $110\, \rm \mu V/m$ (right-hand plot) results in the GRAND frequency band get better than in the full band results for vertical showers only. Eventually, we observe that Radio Morphing reaches its best efficiency in the $[30-80]\, \rm MHz$ band with most standard deviations being even below $10\%$. This highlights that Radio Morphing is more efficient at low frequency. This can be expected as high frequency signals  are located close to Cherenkov cone where the spatial variability of the radio emission is large, which makes the interpolation less efficient. Overall, these results show that Radio Morphing can reasonably reproduce the radio signal from cosmic ray-induced air showers, with relative differences with ZHAireS simulations within the range of calibration uncertainties expected for radio experiments (typically of $20\%$).

\subsection{Timing}

\begin{figure}[tbh]  
    \centering
    \includegraphics[width=\linewidth]{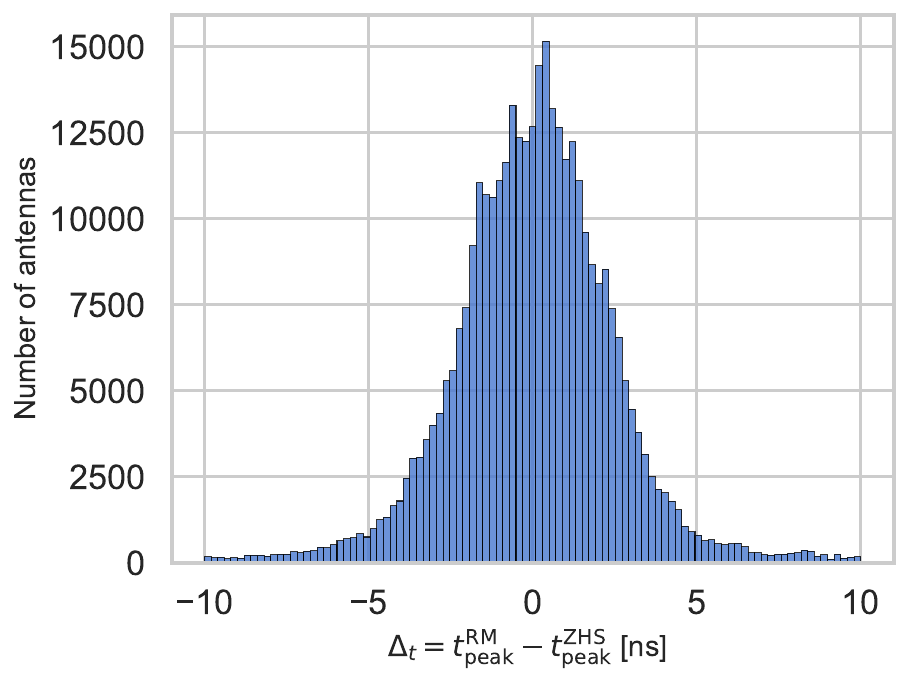}
    \caption{Peak-time difference  at the antenna level between Radio Morphing and ZHAireS simulations. } 
    \label{fig:timedelay}
\end{figure}

We also evaluated Radio Morphing performances on the peak-time reconstruction. A precise timing resolution is essential for the angular reconstruction of the arrival direction of the shower. For this, we evaluated the absolute peak-time difference, at the antenna level, between ZHAireS and Radio Morphing simulations. In the left-hand panel of Fig.~\ref{fig:timedelay}, we show the distribution of the peak-time delay between Radio Morphing and ZHAireS simulations. Radio Morphing outputs are generated with a default sampling rate of $0.5\, \rm ns$. For more than $99\%$ of antennas we get a time difference $\Delta t$ within $\pm 5 {\rm ns}$, i.e., below the typical timing resolution of most radio experiments. This performance is allowed by the new interpolation method used in this version of Radio Morphing~\cite{Zilles2020APh...114...10Z} and is a major improvement compared to the version of~\cite{Chiche2022icrc.confE.194C}. This agreement also suggests that using a straight-line propagation of the radio emission remains a well-motivated approximation even for inclined showers, as was pointed in~\cite{Dieder2025PhRvD.111f3062V} which shows deviation.

\subsection{Computation time}

\begin{figure}[tbh]  
    \centering
    \includegraphics[width=\linewidth]{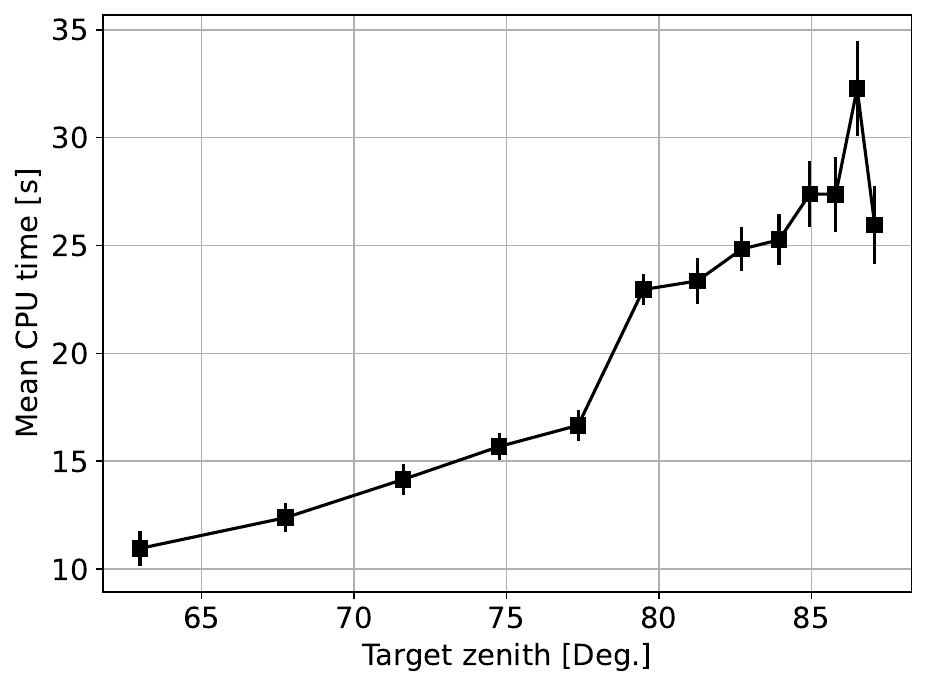}
    \caption{Mean CPU time (square) and standard deviation (error bars) of Radio Morphing simulations as a function of the shower zenith angle. For each shower, the radio emission was simulated at $176$ antenna positions with a thinning of $10^{-5}$.} 
    \label{fig:cputime}
\end{figure}

Eventually, we evaluated the CPU time of Radio Morphing simulations, since the primary objective of the tool is to simulate air shower radio signals for future large-scale radio experiments such as GRAND or the Pierre Auger Observatory. In Fig.~\ref{fig:cputime}, we show the mean CPU time of Radio Morphing as a function of the shower zenith angle from the set of $4500$ showers generated with Radio Morphing to test its performances. Each simulation corresponds to a thinning of $10^{-5}$ and was run for $176$ antennas with a primary energy between $10^{17}\, \rm eV$ and $4\cdot 10^{18}\, \rm eV$. It can be seen that the computation time increases with the zenith angle, which is expected, as more reference planes are used for inclined showers. Still, Radio Morphing provides a computation time of only a few tens of seconds, independently of the shower zenith angle and primary energy.  For a quantitative comparison, 
simulations with the same thinning, primary energy, and number of antennas, run with ZHAireS or CoREAS, are generally expected to take $\sim 1-2$ days of CPU time (with AMD EPYC 7302 16-Core Processor) for $\theta=60^{\circ}$ and around 1 week for $\theta=85^{\circ}$. This leads to an improvement of more than 4 orders of magnitude, making Radio Morphing a suitable tool to run a massive number of simulations in a short amount of time, in preparation for future large-scale radio detectors.

\section{Discussion}

{\it Radio Morphing principle:} The Radio Morphing implementation presented in this article relies on the assumption of a point-like emission at $X_{\rm max}$. This modeling is well justified in the context of a GRAND-like detector, which primarily targets inclined air showers (zenith angle $\theta>60^{\circ}$). For such geometries, the distance between the ground and $X_{\rm max}$ typically reaches tens to hundreds of kilometers, and is therefore much larger than the longitudinal extent of the emitting region, of order a few kilometers. It was hence shown in~\cite{Decoene2023APh...14502779D} that for inclined showers the radio emission peak times could be fitted using a spherical fit. be Radio Morphing is therefore complementary to alternative approaches such as template synthesis~\cite{Mitja2026APh...17503182D} or MGMR~\cite{Scholten2008APh....29...94S, Scholten:2017tcr}, which focus on more vertical showers and can not rely on this approximation. While the point-like assumption simplifies the Radio Morphing implementation, it prevents the modeling of specific features of individual showers, such as double-peaked longitudinal profiles~\cite{Vital2025arXiv251013788D}. Consequently, Radio Morphing is designed to reproduce the average behavior of Monte Carlo simulations, with singular showers being treated as fluctuations around this average.

{\it Scaling procedure:} In the current Radio Morphing method, we assume that the variations of the radio emission with the number of particles at $X_{max}$, the geomagnetic angle, and the air density at $X_{max}$ can be treated independently. We mention that in practice, some coupling may exist between these effects, as some studies for example suggest that variations of the radio emission with the geomagnetic angle can depend on the altitude at $X_{\rm max}$~\cite{chiche2024PhRvL.132w1001C}. Such effects could explain part of the method bias evidence in  the left-hand panel of Fig.~\ref{fig:RMperf_vs_theta} and will be further investigated in future versions of Radio Morphing to improve the method precision.

{\it Reference library and method accuracy:} The reference library used for Radio Morphing was deliberately chosen to use only a few events (16 showers in the current implementation). This choice allows the library to be easily adapted to different configurations, such as changes in the hadronic interaction model or in the geomagnetic field, by re-simulating a restricted set of reference events, which could be done upon demand. We also mention that despite its limited size, the library is densely sampled in zenith angle, enabling the method to capture the main variations of the radio emission with shower inclination by selecting the closest reference shower in our library, without requiring an explicit modeling of complex asymmetry effects~\cite{chiche2024PhRvL.132w1001C, Schlueter2020EPJC...80..643S}. Finally, the precision of the method is tunable and can be adjusted simply by adding more reference showers to better sample the parameter space and reduce the residual differences between the reference and target shower.

{\it Adding neutrino primaries:} We note that while the current Radio Morphing implementation is designed to simulate cosmic rays only, the method could be extended to neutrino primaries using a similar methodology. Although neutrino showers typically develop much closer to the ground than cosmic ray ones, they are expected to follow near-horizontal trajectories  $85^{\circ}<\theta <95^{\circ}$. For such configurations, a kilometer scale ground antenna array would need to be located at least tens of kilometers away from the emission region to have a significant number of antennas ($>5$) intersected by the radio footprint, thus motivating the point-like assumption also in this case.

\section{Conclusion}


We presented the latest version of Radio Morphing, a semi-analytical method that allows for a fast computation of any cosmic-ray induced air shower with zenith angle $\theta>60^{\circ}$,  at any desired position in space, from the simulation data of a few reference ZHAireS showers at given positions. This new version now incorporates shower-to-shower fluctuations, refined scaling laws, interpolation method, and modeling of charge-excess emission. Radio Morphing performances were evaluated by comparison to ZHAireS simulations ran for the same parameters. For most zenith angles, we find mean relative differences in the full band peak amplitude around $5\%$, indicating  only a minor bias of the method, mainly for low-energy showers with an azimuthal direction close to the magnetic North. The corresponding standard deviations on raw traces were evaluated to be below $17\%$ independently of the zenith angle, i.e., below the  calibration error reached by current radio detectors. The standard deviations were even reduced to $15\%$  when applying a $3\sigma$ trigger threshold and below $\sim 10\%$ when working in the $[30-80]\, \rm MHz$ band. These performances were coupled with a timing accuracy of less than  $ 5 \, \rm ns$ and a time computation reduced by more than four orders of magnitude compared to standard Monte Carlo simulations with primary energy above $10^{17}\, \rm eV$. This makes Radio Morphing a powerful and accurate tool for simulating a huge number of air shower simulations in the view of future large-scale radio detectors such as GRAND, SKA, or the radio upgrade of the Pierre Auger Observatory. We note that the current version of Radio Morphing is tuned to the magnetic field configuration of Dunhuang (China). Nevertheless, the method is intrinsically generic and remains applicable to other magnetic field configurations. Extending it to a different site would only require adapting the reference simulations, which could be performed upon request. Further improvements would include the correction of the bias with the shower azimuth angle, an implementation of radio emission scaling with the Earth's magnetic field and the incorporation of other primaries such as neutrinos. Radio Morphing open source code is written in python and now publicly available at \href{https://github.com/simonchiche/RadioMorphing}{https://github.com/simonchiche/RadioMorphing}.


\subsection*{Acknowledgments}

We thank the  CC-IN2P3 Computing Centre (Lyon/Villeurbanne – France), partnership between CNRS/IN2P3 and CEA/DSM/Irfu, for the computing ressources. We thank the GRAND collaborators and in particular Kumiko Kotera for useful discussions and comments.


\bibliographystyle{elsarticle-num} 
\bibliography{biblio}

\end{document}